\documentclass[twocolumn,aps,floatfix,superscriptaddress,longbibliography]{revtex4-1}
\pdfoutput=1 

\usepackage{graphicx,color,enumitem}
\usepackage{dcolumn}
\usepackage{epsfig}

\usepackage[centertags]{amsmath}
\usepackage{amsfonts,amsmath}
\usepackage{euscript}
\usepackage{amssymb}
\usepackage{amsthm}
\usepackage{newlfont}
\usepackage{mathrsfs}
\usepackage{subfigure}
\usepackage{todonotes}
\usepackage{changes}

\usepackage[colorlinks=true, urlcolor=blue, anchorcolor=blue, citecolor=blue,filecolor=blue,linkcolor=blue,menucolor=blue,pagecolor=blue]{hyperref}

\newcommand{\opunit}{\text{1}\kern-0.22em\text{l}}

\newcommand{\eref}[1]{Eq.~(\ref{#1})}

\def\bea{\begin{eqnarray}}
\def\eea{\end{eqnarray}}
\def\ba{\begin{array}}
\def\ea{\end{array}}
\def\n{\nonumber}

\def\la{\langle}
\def\ra{\rangle}

\begin{document}

\title{Universal Dynamics of a Passive Particle Driven by Brownian Motion}
\author{Urna Basu}
\affiliation{S. N. Bose National Centre for Basic Sciences,  Kolkata 700106,  India}
\author{P.  L.  Krapivsky}
\affiliation{Department of Physics,  Boston University,  Boston, Massachusetts 02215,  USA}
\affiliation{Santa Fe Institute, Santa Fe,  New Mexico 87501,  USA}
\author{Satya N. Majumdar}
\affiliation{LPTMS,  CNRS,  Univ.  Paris-Sud,  Universit{\'e} Paris-Saclay,  91405 Orsay,  France}

\begin{abstract}

We investigate the overdamped dynamics of a  `passive' particle driven by 
nonreciprocal interaction with a `driver' Brownian particle. When the interaction between them is short-ranged, the long-time 
behavior of the driven particle is remarkably universal---the mean-squared displacement 
(MSD) and the typical position of the driven particle exhibits the same 
qualitative behaviors independent of the specific form of the potential. In particular, 
the MSD grows as $t^{1/2}$ in one dimension and $\log t$ in two spatial dimensions. We 
compute the exact scaling functions for the position distribution in $d=1$ and $d=2$. These functions are universal when the interaction is short-ranged.  
For long-ranged interactions, the MSD of the driven particle grows as $t^{\phi}$ with 
exponent $\phi$ depending on the tail of the potential.

\end{abstract}
\maketitle

\section{Introduction}

The stochastic dynamics of $N$ interacting particles is a subject of long-standing 
interest in statistical physics. This is a hard problem, except for a few 
solvable cases mostly in one dimension. Even the simple two body problem $(N=2)$ is often 
hard. A two-particle system, with reciprocal interactions between them that depend only on their relative coordinate, can be reduced to a simpler one-body problem in an external potential in their relative co-ordinate. However, when the interaction is non-reciprocal, this simple change of coordinates does not work and the problem is nontrivial even for $N=2$. In this paper, we investigate the stochastic dynamics of two particles with a non-reciprocal interaction between them in all dimensions and derive exact results for the position fluctuations at  asymptototically late times. We find rich and rather universal behaviors that are independent  of the details of the interactions but depend only on the dimension and on whether the nonreciprocal interaction is short-ranged or long-ranged.

Our setup consists of two particles: we call one of them the `driver' particle and the other a `passive' or `driven' particle. We consider a non-reciprocal (one-sided) interaction affecting only the passive particle. We assume that the driver particle performs a standard Brownian motion 
\cite{levy,IM:book,BM:book}, while the motion of the passive particle is overdamped and caused by a central force depending on the separation from the Brownian particle. In one dimension, the position $x(t)$ of the driver particle and the position $y(t)$ of the passive particle evolve according to
\begin{subequations}
\begin{align}
\label{active}
\dot x &= \eta(t),\\
\label{passive}
 \dot y &= f[x(t)-y(t)]\, .
\end{align}
\end{subequations}
Here $\eta(t)$ is a Gaussian white noise obeying standard relations, 
\begin{equation}
\langle \eta(t)\rangle =0, \qquad \langle \eta(t_1) \eta(t_2)\rangle = 2D \delta(t_1-t_2)\, .
\end{equation}

Equations~\eqref{active}-\eqref{passive} provide a simple continuum model  for the motion of a single vacancy in an exclusion process on a lattice. Specifically, a symmetric exclusion process on the square lattice with a single empty site (vacancy) was investigated in Refs.~\cite{Hilhorst,Zoltan}. The vacancy performs a random walk, and whenever it occupies a site neighboring to a tagged particle, the tagged particle can hop, exchanging its position with the vacancy. The driver $x$-particle mimics the motion of this vacancy, which interacts with the tagged particle, modeled by $y$, via a short-ranged interaction.

For a more general scenario, the force acting on the passive particle depends only on the separation between the particles. The force is 
necessarily potential in one dimension, $f(z)=- \partial_z V(z)$. We make a few assumptions about the force, although 
many results remain valid in more general situations. We assume that the force $f(z)$ is an odd function of $z$. 
Equivalently, the potential $V(z)$ is an even function of $z$. However, there is no restriction on the sign of the 
potential, i.e., it can be either attractive or repulsive.

In higher dimensions, we assume that the force is central with the potential $V(\boldsymbol{r}) = V(r)$ that depends 
only on the separation $\boldsymbol {r} = \boldsymbol {x-y}$ between the driver and the driven particles.

The behavior of the position of the driver particle is well-known since it performs standard Brownian motion. The goal 
is to determine the statistics of the position of the passive particle. For instance, we will explore how the 
mean-squared displacement (MSD) grows with time and show that the behavior of the MSD is very different depending on whether the potential is short-ranged or long-ranged. One can thus distinguish between two types of forces $f(r)$ where $r$ represents the distance between the two particles in general dimensions:
\begin{enumerate}
\item {\bf Short-ranged forces}. A short-ranged force $f(r)$ is  finite everywhere and quickly decays with distance, e.g., faster than any power of $1/r$ for $r \gg R$, the typical range of the force.
\item {\bf Long-ranged forces}. A long-ranged force is  finite everywhere, but it decays algebraically, 
\begin{equation}
\label{FA}
f(r)\simeq \frac{A}{r^a} \quad\text{as}\quad r\to\infty, 
\end{equation}
with $a>0$. The force is repulsive if $A>0$, while it is attractive if $A<0$. 
\end{enumerate}

We focus on short-ranged regular forces (no singular behavior as $r\to 0$) and show that the emerging behaviors of the passive particle at late times are universal, e.g., independent of the details of the potential and  whether the force is repulsive or attractive. For long-ranged forces, the behavior may depend on the decay exponent $a$ in \eqref{FA} and also on the sign of the amplitude $A$ in \eqref{FA}, i.e., on whether the force is 
asymptotically repulsive or attractive. For sufficiently large $a$, one anticipates the same behavior as for short-ranged forces. In fact, we show that in one dimension, the potentials \eqref{FA} with $a>3/2$ are effectively short-ranged, viz., the MSD of the passive particle grows in a universal way independent on the decay exponent $a$ and the same as for the short-ranged potentials.

Before going into the details of computations, we first summarize our main results. We start with the case where the force $f(r)$ on the driven particle is short-ranged and characterize its position fluctuations at late times. For simplicity we assume that the driven particle is initially at the origin ($\boldsymbol {y}(0)=0$). The symmetry of the force then ensures that the average position of the driven particle does not change with time, i.e., $\la \boldsymbol{y}(t) \ra = 0$. We find that the effect of the nonreciprocal interaction is captured by the mean-squared displacement (MSD) which shows a remarkable universal behavior. Irrespective of the specific form of the interaction, the MSD of the passive particle increases indefinitely with time in $d\leq 2$ dimensions and saturates when $d>2$:
\begin{equation}
\label{typical}
\sigma_d^2(t) \equiv \la y^2 (t) \ra\sim \left(\frac{V_0 R}{D}\right)^2\times 
\begin{cases}
\sqrt{Dt/R^2}           & d=1\\
\ln(Dt/R^2)              & d=2\\
 O(1)                       & d>2
 \end{cases}
\end{equation}
Here $V_0$ is the typical strength of the short-range potential, $R$ is its range, and $y=|\boldsymbol{y}|$. The dimensionless quantity $V_0/D$ plays the role of the P\'{e}clet number. 

To appreciate the sub-diffusive growth laws \eqref{typical}, let us consider a lattice version in which a driver particle performs a random walk. In one dimension, in the late time limit, the random walker visits $\sim \sqrt{t}$ sites~\cite{Feller,Weiss} and hence it frequently returns to the visited sites and, on an average, spends $\sqrt{t}$ time in each such site. The passive particle is affected when it is close to the random walker. This leads to a heuristic estimate for the typical displacement of the passive particle 
\begin{equation}
\label{1d:est} 
\sum_1^{\sqrt{t}} \pm 1 \sim t^{1/4},
\end{equation}
which, in turn leads to the $\sqrt{t}$ growth of the MSD of the driven particle.

In two dimensions,  the random walker visits $t/\ln t$ sites \cite{Feller,Weiss,Popov21}, so a visited site is visited on average $\ln t$ times. The typical displacement is therefore
\begin{equation}
\label{2d:est}
\sum_1^{\ln t} \pm 1 \sim \sqrt{\ln t},
\end{equation}
leading to the $\log t$ growth of the MSD of the driven particle. For $d>2$, the random walker visits a site a few times and eventually wanders away and never returns \cite{Feller,Weiss}, so the typical spread of the driven particle remains finite. 

\begin{figure}[t]
\includegraphics[width=8.8cm]{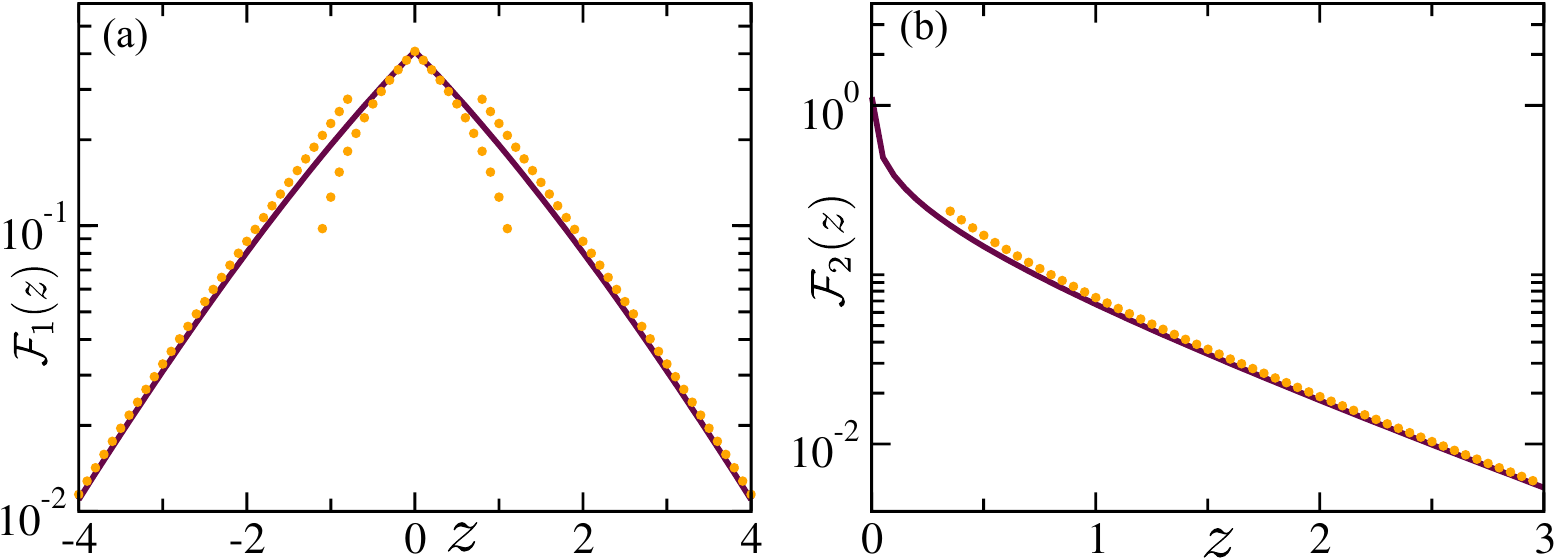} 
\caption{Short-range interaction: Plot of the predicted scaling functions (dark solid lines) in (a) $d=1$ [see Eq.~\eqref{eq:1d_SF}] and (b) $d=2$ [see Eq.~\eqref{eq:2d_SF}]. The dotted lines indicate the asymptotic behaviours quoted in Eqs.~\eqref{1d_asymp} and \eqref{asymp_2d}.} \label{fig:fns_sc}
\end{figure}

It turns out that the complete position distribution of the driven particle is also universal and depends only on the dimension and not on the specific short-ranged potential. In one dimension, the MSD of the passive particle exhibits the same $\sqrt{t}$ growth as that of a tagged particle in single-file diffusion~\cite{Harris_65,Levitt_73,PMR_77,AP_78}. The distribution of the displacement of the tagged particle in single-file diffusion is Gaussian~\cite{A_83}. One could expect the same simple behavior for the displacement of the passive particle, but this is not so. The distribution of the passive particle position $y(t)$ acquires a scaling form 
\begin{subequations}
\begin{equation}
P(y,t) \simeq \frac{1}{c_1 \sigma_1(t)}\, {\cal F}_1(z), \qquad z=\frac{y}{c_1 \sigma_1(t)}
\label{eq:SF_1}
\end{equation}
in the long time limit, with scaling function,
\bea
{\cal F}_1(z)&=& \frac 1{ 2\pi\sqrt{2}} \Bigg[\Gamma \left(\frac{1}{4}\right) \, _0F_2\left(\frac{1}{2},\frac{3}{4};-\frac{z^4}{256}\right)\cr
&& - \sqrt{2 \pi } \, |z|\, _0F_2\left(\frac{3}{4},\frac{5}{4};-\frac{z^4}{256}\right)\cr
&& - \frac{z^2}{2} \Gamma \left(\frac{3}{4}\right) \, _0F_2\left(\frac{5}{4},\frac{3}{2};-\frac{z^4}{256}\right) \Bigg]. 
\label{eq:1d_SF}
\eea
Here 
\bea 
_0F_2(a,b;z) = \sum_{k=0}^\infty \frac{1}{(a)_k (b)_k}\, \frac{z^k}{k!}\,, \label{eq:Fdef} 
\eea
\end{subequations}
is a hypergeometric function where $(a)_k=\frac{\Gamma(a+k)}{\Gamma(a)}$ is the Pochhammer symbol \cite{Knuth}. The scaling function \eqref{eq:1d_SF} is rather complicated but universal, the specific form of the interaction potential only affects the constant $c_1$. The function ${\cal F}_1(z)$ is symmetric in $z$ and has the asymptotic behaviors
\begin{equation}
\label{1d_asymp}
{\cal F}_1(z)\simeq 
\begin{cases}
\displaystyle \frac{1}{2 \sqrt{2} \pi } \Gamma\left[\frac 14\right] - \frac{|z|}{2 \sqrt{\pi}} \quad  & z\to 0  \\[2em]
 \displaystyle \frac {2^{1/6}}{\sqrt{3 \pi}\, z^{1/3}}\, 
\exp\!\left[-3\left(\frac z4 \right)^{4/3}\right] \quad  & z\to \infty .
\end{cases}
\end{equation}
Thus the scaling function has a cusp at $z=0$ and decays super-exponentially as $z\to \infty$. A plot of this scaling function along with the asymptotic behaviours is shown in Fig.~\ref{fig:fns_sc}(a).

In two dimensions, the distribution of the passive particle position is asymptotically isotropic, i.e., depends only 
on $y=|\boldsymbol{y}|$, and has a scaling form
\begin{subequations}
\label{SF_2}
\begin{equation}
P(y,t) \simeq \frac{1}{[c_2 \sigma_2(t)]^2}\, {\cal F}_2(z), \qquad z=\frac{y}{c_2 \sigma_2(t)}.
\label{eq:SF_2}
\end{equation}
The scaling function reads,
\bea
{\cal F}_2(z) = \frac{1}{2\pi}\, K_0(z),
\label{eq:2d_SF}
\eea
\end{subequations}
where $K_0(z)$ is the modified Bessel function of the second kind of order zero. 
The constant $c_2$ is non-universal, viz., it depends on the specific form of the interaction potential. 
The scaling function \eqref{eq:2d_SF} is again universal and, amusingly, simpler than the scaling function 
\eqref{eq:1d_SF} in one dimension. It has the asymptotic behaviors
\begin{equation}
\label{asymp_2d}
{\cal F}_2(z)\simeq 
\begin{cases}
\displaystyle -\frac 1 {2 \pi}\, \ln z+\frac {\ln 2 -\gamma_E}{2 \pi}  \quad & z\to 0 \\[1em]
\displaystyle \frac 1{2\sqrt{2\pi z}}\, \exp(-z) \quad  & z\to \infty,
\end{cases}
\end{equation}
where $\gamma_E = 0.5772 \dots$ is the Euler constant. Thus the scaling function diverges logarithmically as $z\to 0$, while it decays exponentially for large $z$. A plot of the scaling function is shown in Fig.~\ref{fig:fns_sc}(b).

We next investigate the scenario where the passive particle is driven by a long-ranged force of the form \eqref{FA}. We show that in one dimension, the MSD of the driven particle grows algebraically at late times,  
\begin{equation}
\label{1d_MSD_growth}
\la y^2(t)\ra \sim t^\phi \, .
\end{equation}
The exponent $\phi$ is chiefly determined by the decay exponent $a>0$ of the force [cf. Eq.~\eqref{FA}]
\begin{equation}
\label{phi}
\phi = \begin{cases}
\frac{1}{2}     &  a > \frac{3}{2}\\
2-a                & 1< a <\frac{3}{2} \\
\frac{2}{a+1} & 0<a< 1 \quad (\text{repulsive})\\
1                   & 0< a < 1 \quad  (\text{attractive}) \, .
\end{cases}
\end{equation}
The exponent $\phi$ undergoes a freezing transition at $a=\frac{3}{2}$: for all $a>\frac{3}{2}$, the exponent has the same value $\phi=\frac{1}{2}$ as for short-ranged potentials, while for $1< a <\frac{3}{2}$, the exponent depends on $a$ as $\phi=2-a$. In the latter case, the driven particle still moves sub-diffusively but faster than in the case of short-ranged potentials. The scaled position distribution also depends on $a$ in this case. At the transition point $a=3/2$, the growth of the MSD of the driven particle is logarithmically faster than that for short-ranged potentials,
\begin{equation}
\label{y2:log}
\la y^2 (t) \ra\sim \sqrt{t}\,(\log t)^2 \quad  \textrm{for} \quad a=3/2. 
\end{equation}

The behavior in the $a>1$ range remains largely universal, e.g., changing the sign of the potential does not affect the leading asymptotic behavior. However, in the $0<a<1$ range, the nature of the potential plays a role, and qualitatively different behaviors emerge depending on whether the force is repulsive or attractive, i.e.,  whether the amplitude in Eq.~\eqref{FA} is $A>0$  or $A<0$.

The rest of the paper is organized as follows. We start with the case of short-ranged interactions 
(Sec.~\ref{sec:SR}). First, we establish the asymptotic behavior of the MSD in different dimensions 
(Sec.~\ref{sec:MSD}). In Sec.~\ref{sec:Pyt}, we analyze the position distribution employing a heuristic treatment. We 
argue that this method leads to exact results for the scaled position distribution, and support our claims 
numerically. In Sec.~\ref{sec:exact_dist}, we present a few analytical calculations for explicitly tractable 
one-dimensional short-ranged potentials. In Sec.~\ref{sec:long_range}, we study the one-dimensional passive particle 
driven by long-ranged potentials. In Sec.~\ref{sec:concl}, we present concluding remarks. The details of some 
calculations are relegated to the Appendices.

\section{Short-ranged interaction}
\label{sec:SR}

In this section, we explore the position fluctuations of the passive particle when the interaction potential is 
short-ranged. Without loss of generality, we set $x(0)=0$ and $y(0)=0$. Integrating Eq.~\eqref{passive}, we obtain 
\bea
y(t) = \int_0^t d\tau\,f[x(\tau)-y(\tau)]  
\label{eq:yt}
\eea
For simplicity, we use the one-dimensional notation; it is straightforward to generalize to higher dimensions [see Appendix \ref{ap:msd} for details]. For short-ranged potentials, we anticipate that $|x(t)| \gg |y(t)|$. Hence, Eq.~\eqref{eq:yt} reduces to a Brownian functional,
\bea
y(t)\simeq \int_0^t d\tau \,  f[x(\tau)].  \label{eq:yt_approx}
\eea
Brownian functionals appear in various contexts both in physics and mathematics, and they are well studied~\cite{BF_2005,BF23}. The long-time approximation of the driven particle position being given by a Brownian functional allows us to investigate the nature of fluctuations of $y(t)$. 

\begin{figure}
\includegraphics[width=7 cm]{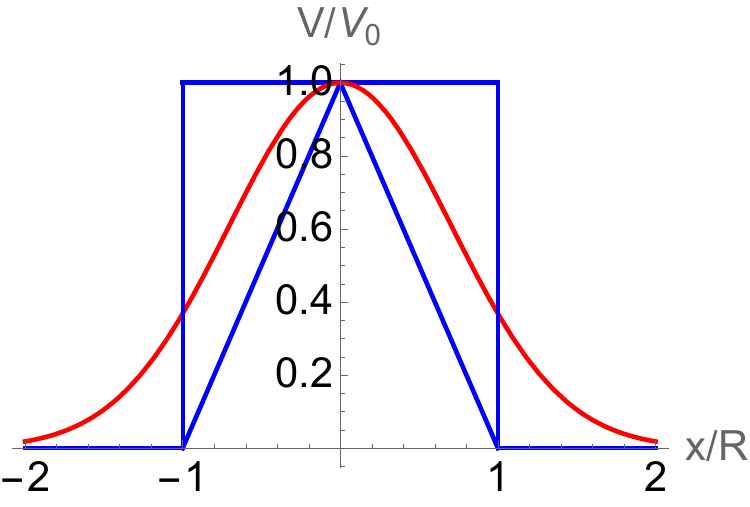}
\caption{The square, Gaussian, and tent potentials in one dimension (top to bottom in the $|x|<R$ range). }
\label{fig:potentials}
\end{figure}

\subsection{Mean-squared displacement}
\label{sec:MSD}

We start with computation of the second moment $\la y^2(t)\ra$, which is the same as the MSD since $\la y(t) \ra =0$ due to our conventions about the intial condition. We are interested in the long time behavior, so we employ Eq.~\eqref{eq:yt_approx} and obtain
\bea
\la y^2(t)\ra =2 \int_0^t dt_2 \int_0^{t_2} d t_1\, \Big \la f(x(t_2)) f(x(t_1)) \Big \ra, \label{eq:y2_def}
\eea
where we have used the symmetry of the two-point correlation of the force $\la f(x(t_1)) f(x(t_2))\ra $ under the exchange of $t_1$ and $t_2$.  The two-point correlation, for $t_2 > t_1$, reads
\begin{eqnarray}
\Big \la f(x(t_2)) f(x(t_1)) \Big \ra & = &  \int_{-\infty}^\infty dx_2  \int_{-\infty}^\infty dx_1 \, f(x_2) f(x_1) \nonumber \\
& \times & \mathcal {P}(\boldsymbol{2}| \boldsymbol{1}) \mathcal{P}(\boldsymbol{1}| \boldsymbol{0}). 
\label{eq:fcor_def}
\end{eqnarray}
We use the shorthand notations {\bf 2}=$(x_2,t_2)$,  {\bf 1}=$(x_1,t_1)$, and {\bf 0}=$(0,0)$. We denote by $\mathcal{P}(\boldsymbol{2}| \boldsymbol{1})$ the probability density for the Brownian particle to move from $\boldsymbol{1}$ to $\boldsymbol{2}$; similarly, $\mathcal{P}(\boldsymbol{1}| \boldsymbol{0})$ is the probability density to move from $\boldsymbol{0}$ to $\boldsymbol{1}$.  These propagators are Gaussian; e.g., 
\begin{align}
 \mathcal {P}(\boldsymbol{2}| \boldsymbol{1}) = \frac 1{\sqrt{4 \pi D (t_2-t_1)}}\,\exp\!{\left[- \frac {(x_2-x_1)^2}{4 D (t_2-t_1)}\right]}\,.
\end{align}
To compute the two-point correlation \eqref{eq:fcor_def}, it is convenient to express the above propagator in Fourier space and substitute it in \eref{eq:fcor_def}. Then, performing the $x$-integrals, we get [see Appendix~\ref{ap:msd} for the details]
\bea
\Big \la f(x(t_1)) f(x(t_2)) \Big \ra &=& \int \frac{dk_1}{2 \pi} \int \frac{dk_2}{2 \pi} \tilde f(k_1 - k_2 ) \tilde f(k_2)\cr
&\times & \exp {[-D (k_1^2 t_2+k_2^2 (t_1 -t_2)]},\qquad \label{eq:f12}
\eea
where $\tilde f(k) = \int_{-\infty}^\infty dz\, e^{i kz} f(z)$ is the Fourier transform of the force $f(z)$. Substituting \eqref{eq:f12} in \eqref{eq:y2_def}, and performing the integrals over $t_1$ and $t_2$, we get
\bea
\la y^2(t) \ra &=& \frac 2{D^2} \int_{-\infty}^\infty \frac{dk_1}{2 \pi} \int_{-\infty}^{\infty} \frac{dk_2}{2 \pi} \frac{\tilde f(k_1 - k_2) \tilde f(k_2)}{k_1^2 k_2^2(k_1^2-k_2^2)}\cr
&& \times  \left[k_1^2 (1- e ^{-D k_2^2 t})-k_2^2(1- e^{-D k_1^2 t})\right].~~\label{eq:y2_kint} 
\eea
The above expression gives the exact MSD for the process \eqref{eq:yt_approx}, but it cannot be evaluated without knowing the explicit form of the force. However, we are primarily interested in the long-time behavior where the process \eqref{eq:yt_approx} provides an asymptotically exact description of the actual evolution of the passive particle. Therefore, we focus on the large-time behavior of Eq.~\eqref{eq:y2_kint} where we can advance further. To this end, let us consider the Laplace transform of the second moment, $M_2(s) = \int_0^\infty dt\, e^{-st} \la y^2(t) \ra$. From Eq.~\eqref{eq:y2_kint} we then have
\bea
M_2(s) &=& \frac 2s \int_{-\infty}^{\infty} \frac{dk_1}{2 \pi} \int_{-\infty}^{\infty} \frac{dk_2}{2 \pi} \frac{\tilde f(k_1 - k_2) \tilde f(k_2)}{(D k_1^2+s)(Dk_2^2+s)}.\qquad \label{eq:y2_laplace}
\eea
The large-time behavior of the MSD is controlled by the small $s$ behavior of its Laplace transform $M_2(s)$. The small $s$ behavior of the Laplace transform $M_2(s)$ can be extracted systematically from Eq.~\eqref{eq:y2_laplace}, and it turns out that it is independent of the specific form of $\tilde f(k)$. See Appendix~\ref{ap:msd} for the details of the computation. Here, we only sketch the main steps.  

For small $s$, the dominant contribution to the integrals in Eq.~\eqref{eq:y2_laplace} comes from the region $-\Lambda \le (k_1,k_2) \le \Lambda$ where $\Lambda \sim O(1)$ is an ultraviolet cut-off. Consequently, it suffices to consider the small $k$ behavior of $\tilde f(k)$, which is analytic for a short-ranged force.  Hence, for small $k$, we must have, $\tilde f(k) = -i a_1 k + O(k^2)$ where $a_1 = 2 \int_0^\infty dz \, z V'(z)$ contains the information about the specific interaction potential. Using this approximation and performing the integrals in Eq.~\eqref{eq:y2_laplace}, we get
\bea
M_2(s) \simeq \frac {a_1^2 \Lambda }{\pi  (Ds)^{3/2}}
\label{eq:y2_smalls_1d}
\eea
for small $s$ [see Appendix~\ref{ap:msd} for the details]. The above equation immediately leads to
\bea
\la y^2(t) \ra  \simeq \frac{2 a_1^2 \Lambda}{(\pi D)^{3/2}}\, \sqrt{t}
\eea
for short-ranged forces in one dimension. Clearly, the specific form of the potential only affects the numerical factor $a_1$, and the asymptotic $\sqrt{t}$ growth of the MSD of the driven particle is universal for short-ranged interactions. This sub-diffusive growth law is one of the main results of this paper.

It is straightforward to generalize the above analysis to higher dimensions. The $k_1$ and $k_2$ integrals in Eq.~\eqref{eq:y2_laplace} become $d$-dimensional [see Eq.~\eqref{eq:y2_laplace_d}]. For small $s$, the dominating contribution to the $k$-integrals comes again from the small $|k|$ regime. For short-ranged rotationally symmetric interactions in two dimensions, we get $ \tilde {\boldsymbol{f}}(\boldsymbol k) = -ia_2 {\boldsymbol k} + O(k^2)$ where $a_2=\frac \pi 2 \int_0^\infty dz\, z^2 V'(z)$ contains the information about the interaction. Using the linear approximation for $ \tilde {\boldsymbol{f}}(\boldsymbol k)$, the $k$ integrals can be performed [see Appendix~\ref{ap:msd}]. Expanding the resulting expression for small $s$ we arrive at 
\bea
M_2(s) \simeq  \frac{a_2^2}{8 \pi^2 D^2} \frac{\log (D/s)}{s}, \label{eq:small_s_d2}
\eea
which can be inverted to obtain the long-time behavior of the MSD in two dimensions:
\bea
\la y^2(t) \ra \simeq \frac{a_2^2}{8 \pi^2 D^2} (\gamma_E + \log Dt), 
\eea
where $\gamma_E$ denotes the Euler constant. Once again, we find a universal behavior, namely a $\log t$ growth of the MSD, in $d=2$.

Following the same analysis, in $d=3$ we obtain
\bea
M_2(s) \simeq  \frac{a_3^2 \Lambda^4}{6 \pi^4 D^2 s} + O(s^{-3/2}), \label{eq:m2s_3d}
\eea
which, in turn, implies that the MSD eventually reaches a constant value in three dimensions. The same holds for all $d\geq 3$. 

\begin{figure}[t]
\includegraphics[width=8.8cm]{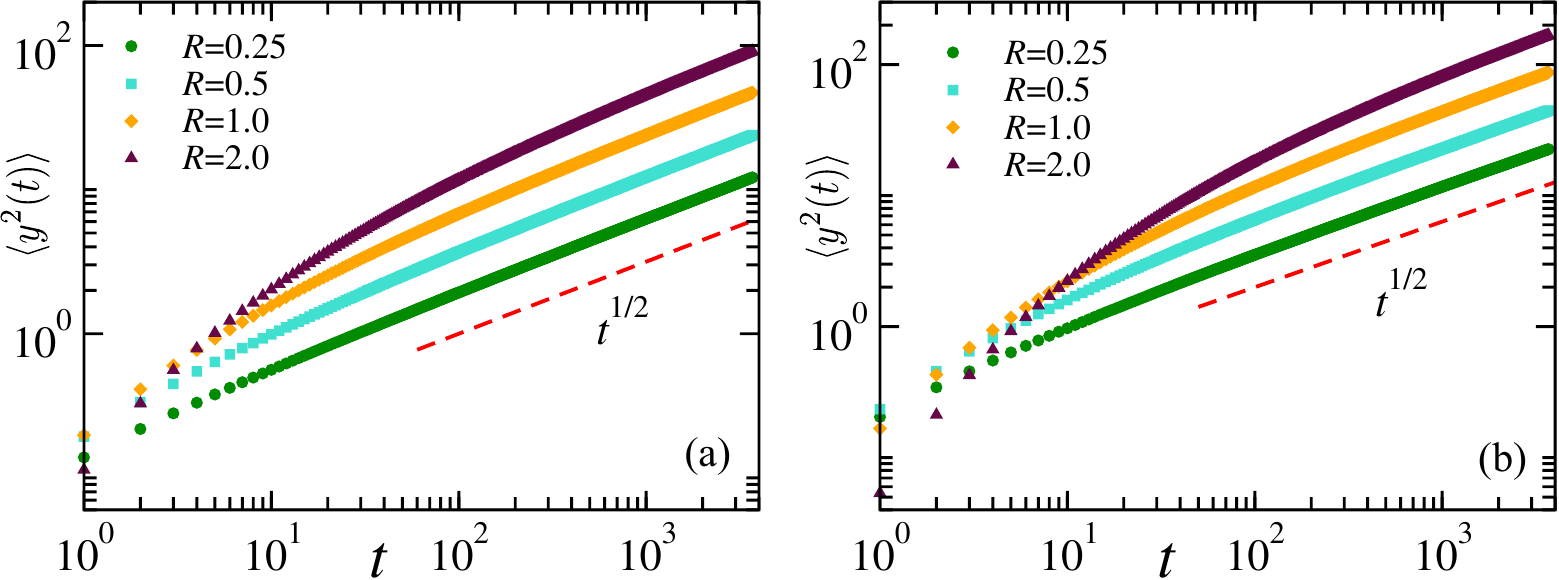} 
\caption{The MSD of the driven particle for short-ranged interaction in $d=1$: Plot of $\la y^2(t)\ra$ versus $t$ measured from numerical simulations using (a) tent potential [see Eq.~\eqref{eq:V_lin}] and (b) Gaussian potential [see Eq.~\eqref{eq:V_gauss}]. Different curves correspond to different values of the range of interaction $R$, and the red dashed curves indicate the analytically predicted $\sqrt{t}$ behavior. Here, we have taken  $V_0=1$ and $D=1$.}\label{fig:yvar_1d}
\end{figure}

We illustrate the universal behavior  of the MSD at late times in one and two dimensions, using numerical simulations for two different short-ranged interaction potentials. First is a tent potential, which is nothing but a truncated linear potential,
\bea
V(\boldsymbol{r}) = 
\begin{cases}
 V_0 \Big(1- \frac{r}{R} \Big) ~\text{for}~ r<R \cr
 0 \qquad \qquad \quad \text{otherwise}.
\end{cases} \label{eq:V_lin}
\eea
The driven particle in the tent potential feels a constant force $V_0/R$ whenever its distance from the Brownian particle is less than $R$, the range of the  potential \eqref{eq:V_lin}. Our second example is a Gaussian interaction potential,
\bea
V(\boldsymbol{r}) = V_0\,e^{-(r/R)^2}, 
\label{eq:V_gauss}
\eea
which has a typical range $R$. The potentials \eqref{eq:V_lin} and \eqref{eq:V_gauss} are attractive when $V_0<0$ and repulsive when $V_0>0$. In one dimension, we performed exact analytical calculations for the tent and Gaussian potential, and for the square potential (Sec.~\ref{sec:exact_dist} and Appendixes). Schematic representations of these potentials are shown in Fig.~\ref{fig:potentials}.

\begin{figure}[t]
\includegraphics[width=8.8 cm]{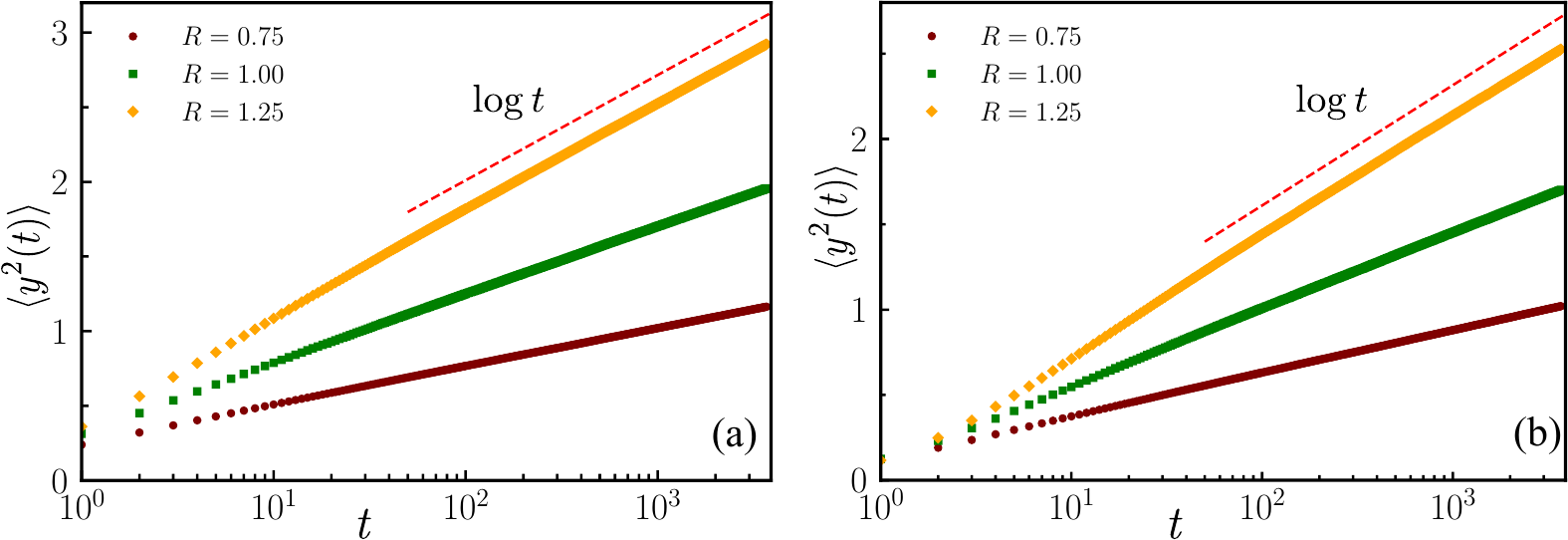}
\caption{The behavior of the MSD of the driven particle in $d=2$ for (a) tent potential [see Eq.~\eqref{eq:V_lin}] and (b) Gaussian potential [see Eq.~\eqref{eq:V_gauss}], obtained from numerical simulations with $D=1$ and the same $V_0=1$ for both potentials and three different values of the range $R$. We have taken $x_0=0.1$ for (a) and $x_0=0$ for (b). The red dashed lines on both panels indicate the $\log t$ behavior. }
\label{fig:yvar_2d}
\end{figure}

We performed numerical simulations of the two-particle dynamics \eqref{active}--\eqref{passive} using interaction potentials \eqref{eq:V_lin} and \eqref{eq:V_gauss} in one and two spatial dimensions.  Figure~\ref{fig:yvar_1d} illustrates that, for $d=1$, the MSD asymptotically grows as $\sqrt{t}$ for both interaction potentials. The results from the $d=2$ simulations are shown in Fig.~\ref{fig:yvar_2d} which illustrate the logarithmic growth, $\la y^2(t) \ra \sim \log t$, for both interaction potentials.

\subsection{Position distribution}
\label{sec:Pyt}

The universal nature of the MSD of the driven particle raises the obvious question of whether similar universal features are manifest in the full position distribution of the driven particle. To answer this question,  we investigate the position distribution $P(y,t)$ at large-times $t$ using two different approaches. Both approaches apply in the long time limit and give exact scaled distributions. Their status is different, however. A more rigorous approach based on the Feynman-Kac formalism is challenging, and we implemented it only in some specific examples of short-ranged potentials in one dimension where we succeeded in computing the distribution explicitly (Sec.~\ref{sec:exact_dist}). The second approach is less rigorous, and we call it a heuristic argument. This approach relies on the first-passage properties of the Brownian particle, and its obvious advantage is that the computations do not depend on the specific form of the 
short-ranged potential that affects only a non-universal constant. We implemented the second approach both in one and two dimensions. Whenever both approaches are applicable, they give the same universal scaling behavior. Here, we show how to determine the scaling functions in the realm of the heuristic approach that is less rigorous but applies to arbitrary short-ranged potentials in one and two dimensions.

The position distribution $P(y,t)$ of the driven particle is affected when it is close to the position $x(t)$ of the Brownian particle. Hence, the displacement of the driven particle can be estimated as
\bea
y = \sum_{j=1}^N \eta_j,  \label{eq:ysum} 
\eea
where $N$ denotes the number of times the Brownian particle comes close to the driven particle, and $\eta_j$ denotes the displacement of the driven particle at each such instance. We have already used a simpler version of the above equation, viz. Eq.~\eqref{1d:est}, for a heuristic estimate of the typical displacement of the driven particle. Equation \eqref{eq:ysum} also looks like a heuristic estimate. However, the discrete formulation implied by Eq.~\eqref{eq:ysum} becomes asymptotically exact, and the instances when the interaction is non-negligible become independent. Indeed, the driven particle is much `slower' than the Brownian particle, so it does not move far from its initial position. Hence, one can think about $N$ as the number of returns of the Brownian particle to the origin. The potential must be 
short-ranged to justify the discrete treatment of interaction instances. Thus the goal is to improve the naive estimate in Eq.~\eqref{1d:est} where we replaced $N$ by a typical number of return events. In  Eq.~\eqref{1d:est}, we also used the simplest estimate $\eta_j=\pm 1$. This simplification only misses the non-universal constant $c_1$, but we can restore it at the end using the calculation of the MSD presented in Sec.~\ref{sec:MSD}. 

The above arguments imply that position distribution of the driven particle can then be expressed as
\bea
P(y,t) = \sum_{N} P(y|N) P(N|t) \label{eq:Pyt_def}
\eea
where $P(y|N)$ denotes the probability that the driven particle has a displacement $y$, given $N$ return events, and $P(N|t)$ denotes the conditional probability that there are $N$ returns of the Brownian particle to the origin, during the interval $[0,t]$.  Note that, for the sake of simplicity, we have used the same letter $P$ to denote all probability distributions. 

To compute $P(N|t)$ we adopt a trajectory based approach. Let us consider a trajectory of the Brownian particle with $N$ returns to the origin and let $\tau_i$ denote the interval between $i-1$-th and $i$-th returns. The probability of such a trajectory is 
\bea
P(N, \{\tau_i \}|t)= \prod_{i=1}^{N} g(\tau_i) q(\tau_{N+1}) \delta( t - \sum_{i=1}^{N+1} \tau_i)
\eea
where $g(\tau)$ denotes the first passage probability of the Brownian motion and
\bea 
 q(\tau_{N+1})= \int_{\tau_{N+1}}^{\infty} d\tau' \, g(\tau')  \label{eq:qt_def}
\eea 
denotes the probability that there were no returns during the final interval $\tau_{N+1}$. Combining the contributions from all such trajectories with $N$ returns we arrive at 
\begin{align}
P(N|t) = \int \prod_{i=1}^{N} d \tau_i \, g(\tau_i) \, d\tau_{N+1} \, q(\tau_{N+1})\, \delta\Big(t - \sum_{i=1}^{N+1} \tau_i\Big).\label{eq:PNt_def}
\end{align}
It is convenient to consider its Laplace transform,
\bea
\tilde P(N,s) = \int_0^{\infty} P(N|t) e^{-st} dt = [\tilde{g}(s)]^{N} \tilde{q}(s), \label{eq:PNs}
\eea
where $\tilde g(s)$ and $\tilde q(s)$ denote the Laplace transforms of $g(t)$ and $q(t)$, respectively.  These two quantities are related. Indeed, performing the Laplace transform of \eref{eq:qt_def} and using $q(0)=1$ we obtain
\begin{align}
\label{qs}
\tilde q(s) = \frac{1-\tilde g(s)}{s}\,.
\end{align}
In the following, we use \eref{eq:PNt_def} along with the known results for $P(y|N)$ to compute $P(y,t)$ in $d=1$ and $d=2$. \\

\begin{figure}[t]
\includegraphics[width=7 cm]{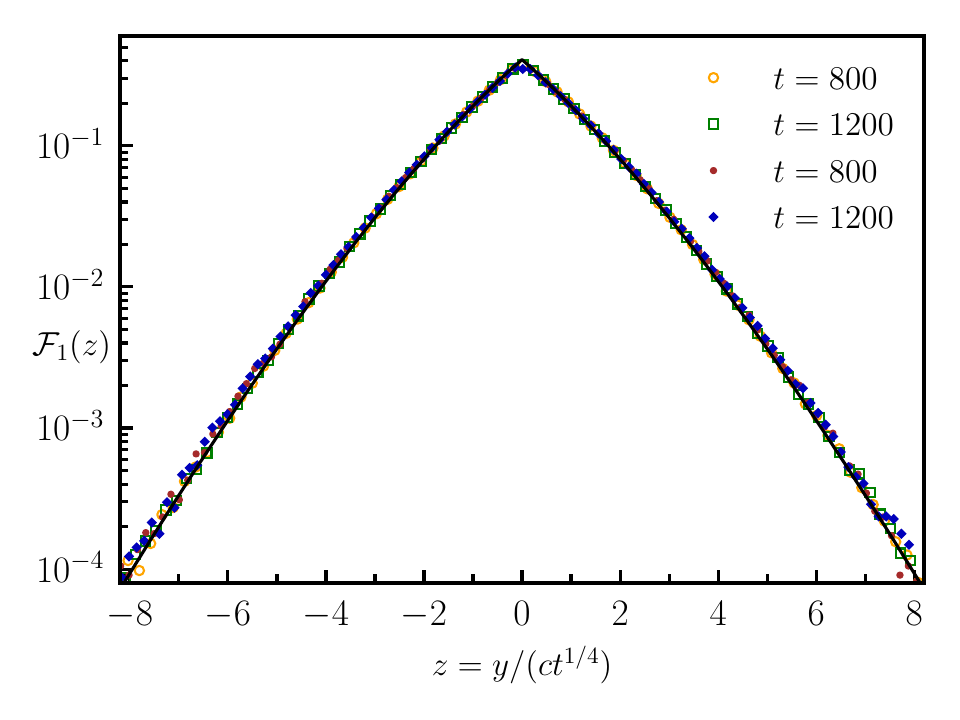}
\caption{Scaled position distribution of the driven particle in 
$d=1$ measured from numerical simulations with tent potential (open symbols) and Gaussian potential (filled symbols). 
We have taken $D=1,R=0.5$, $V_0=1$, and $x_0=0$. The solid black line indicates the analytically 
predicted scaling function~\eqref{eq:1d_SF}. The non-universal scaling factor, $c$ in the figure, 
is different for the two interaction potentials.}
\label{fig:ydist_1d}
\end{figure}

In one spatial dimension, the driven particle can only move in two directions. The displacements, which can be set to unity as we argued before as along as we are interested only in the functional form of the scaled position distribution, are therefore $\eta_i=\pm 1$. The distribution of $y$ for a given number, $N$, of interactions with the Brownian particle is asymptotically Gaussian:
\bea
P(y|N)= \frac 1{\sqrt{4 \pi N}} \exp{\left(-\frac{y^2}{4 N}\right)},
\eea
for $N\gg 1$. Next, we need to compute $\tilde P(N,s)$ which requires  determination of $\tilde g(s)$ [cf. Eqs.~\eqref{eq:PNs} and \eqref{qs}]. The first passage properties of one-dimensional Brownian motion are well studied and the exact form of $g(\tau)$ is known \cite{Weiss},
\bea
g(\tau) = \frac{|\Delta x|}{\sqrt{4 \pi D \tau^3}} \exp{\left [- \frac{(\Delta x)^2}{4 D \tau} \right]}, 
\eea
where $\Delta x$ denotes the initial distance of the Brownian and driven particles. The corresponding  Laplace transform is $\tilde g(s) = e^{- b \sqrt{s}}$ with 
$b= \Delta x/\sqrt{D}$. We can set $b=1$ as the precise value only affects the non-universal constant $c_1$ which we fix otherwise [see Sec.~\ref{sec:MSD} and Appendix~\ref{ap:msd}]. Using $\tilde g(s) = e^{-\sqrt{s}}$ and Eq.~\eqref{qs} we have $\tilde q(s) = s^{-1}\big(1- e^{-\sqrt{s}}\big)$ and deduce 
\begin{equation}
\tilde P(N,s) = \frac{1- e^{-\sqrt{s}}}{s}\,e^{-N\sqrt{s}}
\end{equation}
from Eq.~\eqref{eq:PNs}. We are primarily interested in the large $t$ behavior, hence, it suffices to consider the small $s$ behavior of the Laplace transform, 
\bea
\tilde P(N,s) \simeq \frac{e^{-N\sqrt{s}}}{\sqrt{s}}\,, 
\eea
which we invert to find
\bea
P(N,t) \simeq \frac 1{\sqrt{\pi t}} \exp{\left(-\frac{N^2}{4 t}\right)}.
\eea
Replacing the summation in Eq.~\eqref{eq:Pyt_def} by integration we obtain
\bea
\label{pdist.1}
P(y,t) \simeq \int_0^{\infty} \frac{dN}{\sqrt{4 N \pi^2 t}}\, \exp\!\left[-\frac{y^2}{4 N}- \frac{N^2}{4t}\right]. 
\eea
Let us remark that this heuristic argument esentially maps the effective motion of the passive particle to a continuous-time random walk (CTRW) problem~\cite{MS1973}. CTRW refers to a walker that performs  $\pm 1$ random walk of $N$ steps,
but before every jump the walker waits at a lattice site for a random time $\tau$
drawn, independently for each jump, from a waiting time distribution $g(\tau)$.

The distribution \eqref{pdist.1} can be expressed in a  scaling form
\bea
P(y,t) \simeq \frac 1{t^{1/4}} {\cal F}_1\left(\frac y {t^{1/4}} \right)
\eea 
where the scaling function is given by
\bea
{\cal F}_1(z) = \frac 1{\pi \sqrt{2}} \int_0^\infty 
\frac {du}{\sqrt{u}}\, \exp\!\left[- \frac{z^2}{8u}- u^2 \right]\, . 
\label{eq:1d_int_SF}
\eea
The integral can be evaluated exactly leading to the announced scaling function \eqref{eq:1d_SF}. 
This is one of the central results of this work---for short-ranged interactions, the distribution of typical position 
fluctuations of the driven particle in the large-time regime is universal and independent of the specific form of the 
interaction potential. The asymptotic behaviors of the scaling function $\cal{F}_1(z)$ are given in \eqref{1d_asymp}.
Note that the large $z$ behavior in \eqref{1d_asymp} is easier to extract by performing a saddle point
analysis of the integral representation \eqref{eq:1d_int_SF}, rather than from the formal expression
in \eqref{eq:1d_SF}.

The universal nature of the position fluctuations of the driven particle is illustrated in Fig.~\ref{fig:ydist_1d} 
comparing the scaled position distribution obtained from numerical simulations with interaction potential 
\eqref{eq:V_lin} and \eqref{eq:V_gauss} with the analytical prediction Eq.~\eqref{eq:1d_SF}. The match is excellent. 
We used a numerical factor to obtain the collapse for the two different interaction potentials. This non-universal 
factor $c_1$ appearing in Eq.~\eqref{eq:SF_1} depends on the specific form of the interaction and cannot be determined 
from the heuristic argument presented above. In Sec.~\ref{sec:exact_dist}, we discuss some specific examples where we 
deduce this factor from an exact computation of the distribution $P(y,t)$. \\

\begin{figure}
\includegraphics[width=8.8 cm]{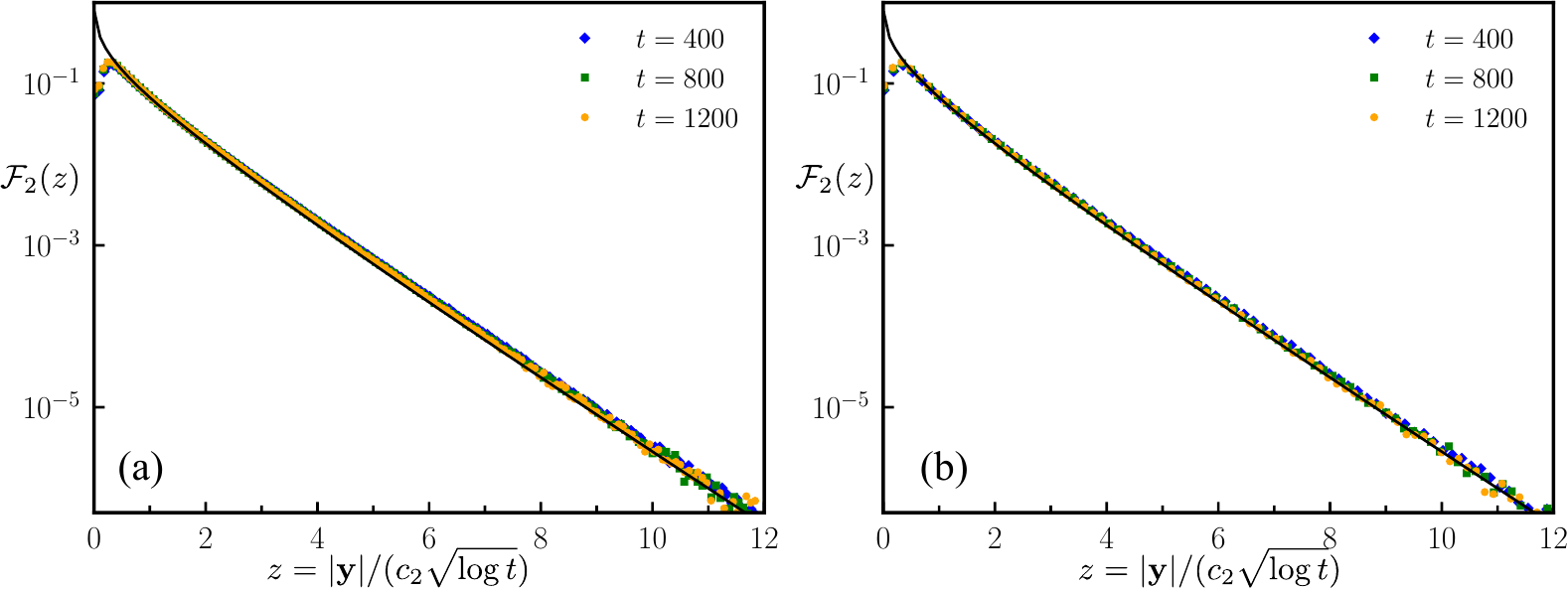}
\caption{Scaled position distribution of the driven particle in $d=2$ for (a) tent potential and (b) Gaussian potential. The different curves correspond to the distribution of the scaled position $z= |y|/ (c_2 \sqrt{\log t})$ obtained from numerical simulations; the non-universal factor $c_2$ is different for the two potentials. The solid black lines correspond to the analytical prediction~\eqref{eq:2d_SF}. We have taken $D=1,V_0=1$ and $R=0.5$ for both the cases.}
\label{fig:2d_dist}
\end{figure}

This heuristic argument can be extended to two dimensions. In this case, the approximation \eqref{eq:ysum}  describes a two-dimensional random walker implying a Gaussian distribution of $y$ for a given $N$:
\bea
P(y|N)= \frac 1 {4 \pi N}\,  \exp\!\left[-\frac{|y|^2}{4N}\right].\label{eq:PyN_2d}
\eea
The next step is to obtain $P(N|t)$, i.e., the probability that the two-dimensional Brownian particle $x$ has $N$ return events to the origin within a duration $t$,  which is given by \eref{eq:PNt_def}. A convenient way to find $P(N|t)$ is again via its Laplace transform [see \eref{eq:PNs}]. Thus, we need $g(s)$,  the Laplace transform of the first-passage time distribution.  For a two-dimensional Brownian motion, the first-passage time is the time to return within a radial distance $a$ of the origin, starting from a radial distance $b>a$.  Let $S(t,a,b)$ be the corresponding survival probability, i.e., the probability that, starting from a distance $b$, the Brownian particle has not returned within a distance $a$ of the origin up to time $t$. It is known that the Laplace transform of $S(t,a,b)$ is given by~ \cite{Spitzer},
\bea
\int_0^t dt \, e^{-st} S(t,a,b) = \frac{K_0(b \sqrt{2s})}{s K_0(a \sqrt{2s})}\,,
\eea
where $K_0(z)$ is the modified Bessel function of the second kind of order zero. The corresponding first-passage time distribution is given by $g(t) = - \partial_t S(t,a,b)$, so its Laplace transform reads
\bea
\tilde g(s) = \frac{K_0(b \sqrt{2s})}{K_0(a \sqrt{2s})}\,.
\eea
Since we are interested in the long-time regime, it suffices to consider the small $s$ behavior of $\tilde g(s)$,
\bea
\tilde g(s) = 1+ \frac{2 \log(b/a)}{\log s} +O(s \log s).
\eea
Consequently
\bea
\tilde q(s) \simeq - \frac{c}{s \log s},
\eea
where $c=2 \log(b/a)$ is a non-universal constant. Using this result together with Eqs.~\eqref{eq:PNs}--\eqref{qs} we can write
\bea
\tilde P(N,s) \simeq   \frac{c}{s \log s} \exp{\bigg(\frac {cN}{\log s}\bigg)}
\eea 
for small $s$ and large $N$. From the above equation, one can get the large time behavior of $P(N,t)$ by inverting the Laplace transform using Bronwich integral, 
\bea
P(N,t) = - \frac{c}{2 \pi i} \int_{\Gamma} \frac{ds}{s \log s} e^{st}  \exp{\bigg(\frac {cN}{\log s}\bigg)}
\eea
where $\Gamma$ denotes the standard Bromwich contour which encloses all the singularties of the integrand. It is convenient to use a change of variable $z=st$ which yields
\bea
P(N,t) =- \frac{c}{2 \pi i} \int_{\Gamma} \frac{dz}{z(\log z - \log t)} e^z \exp{\bigg(\frac {cN}{\log  z - \log t}\bigg)}.~ \n
\eea 
This integral is dominated by the region where $z=O(1)$, i.e., $\log z \ll \log t$. Thus 
\bea
P(N,t) \simeq \frac c {\log t} \exp \left(- \frac{c N}{\log t} \right)  \frac 1 {2\pi i} \int_{\Gamma} \frac {dz}z e^z
\eea
from which 
\bea
\label{PNt:2}
P(N|t) \simeq \frac 1{\log t} \exp{\left(-\frac{N}{\log t} \right)},
\eea
where, once again, we have put the non-universal constant $c=1$. 

Substituting \eqref{eq:PyN_2d} and \eqref{PNt:2} into Eq.~\eqref{eq:Pyt_def} and replacing summation by integration we arrive at
\bea
P(y,t) & \simeq & \frac 1{4 \pi \log t} \int_0^{\infty} \frac{dN}{ N } \exp{\Big(- \frac{y^2}{4N}- \frac{N}{\ln t}\Big)} \cr
&=& \frac 1{\ln t} {\cal F}_2 \left( \frac{y}{\sqrt{\ln t}}\right)
\eea
with the scaling function given by
\bea
\label{Bessel}
{\cal F}_2(z)= \frac 1{4\pi} \int_0^{\infty} \frac {du}{u}\, \exp\!\left[-\frac{z^2}{4u}- u \right]= \frac{1}{2\pi}\, K_0(z)\, ,  
\eea
as announced in \eqref{eq:2d_SF}.

Figure \ref{fig:2d_dist} shows a comparison of the distribution of the scaled radial distance $z=|y|/(c_2 \sqrt{\log t})$ of the driven particle, obtained from numerical simulations for interaction potentials \eqref{eq:V_lin} and \eqref{eq:V_gauss}, with the predicted scaling function \eqref{eq:2d_SF}. By adjusting the non-universal factor $c_2$ an excellent match is observed in both cases. The only exception is the small $z$ behavior where the ultimate logarithmic divergence [cf. Eq.~\eqref{asymp_2d}] is difficult to see at a large but finite time. 

\begin{figure}
\includegraphics[width=7 cm]{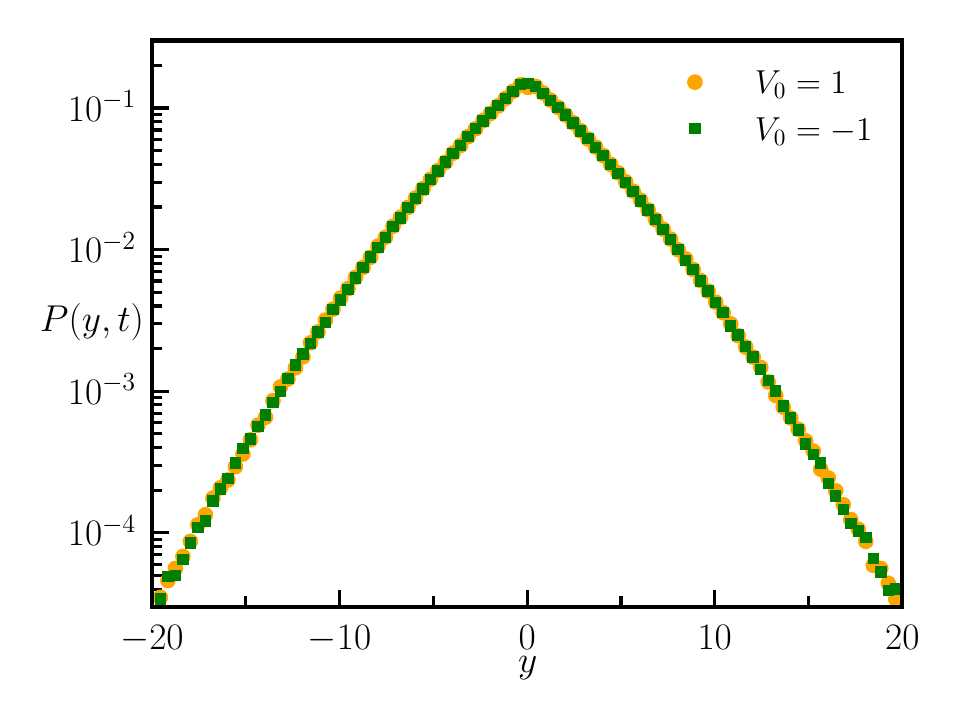}
\caption{Plot of $P(y,t)$ obtained from numerical simulations in $d=1$ with repulsive ($V_0=1$) and attractive $(V_0=-1)$ Gaussian interaction potential \eqref{eq:V_gauss}.  We have taken $t=400$, $R=0.5$ and $D=1$.}
\label{fig:ydist_sign}
\end{figure}

The slow logarithmic growth of the MSD of the driven particle in our continuum model in two dimensions reproduces a similar anomalous growth of a tagged particle on a two dimensional lattice mediated by a single vacancy~\cite{Hilhorst,Zoltan}.
Moreover, even the scaled position distribution of the tagged particle is recovered in Eq.~\eqref{eq:2d_SF} in our simpler continuum model. The analysis of the lattice version in Refs.~\cite{Hilhorst,Zoltan} is much more complicated than our case. Our analysis also has the advantage of leading to universal results independent of the short-ranged interaction potential, and even on the nature of the potential, i.e., whether it is attractive, repulsive, or a  combination of both. An interesting attempt~\cite{Newman} of developing a continuum theory of vacancy-mediated diffusion relied on a mean-field approximation, and was also aimed at modeling the original lattice problem~\cite{Hilhorst,Zoltan}. Our setting is continuous from the beginning, and we also treated arbitrary short-ranged interaction potentials. As often happens in exact sciences, a more general framework is frequently more tractable than a specific one.

For the one-dimensional symmetric exclusion process with a single empty site, the motion of the tagged particle is trivial---it first hops to the right (assuming that the vacancy was originally to the right of the tagged particle), then to the left, then to the right, etc. Thus in this case, the tagged particle is essentially confined only to two sites. A non-trivial vacancy-meditated diffusion is feasible on quasi-one-dimensional lattices, e.g., on the ladders. The emerging behavior is essentially the same as the behavior of our driven particle in one dimension. The analysis~\cite{Hilhorst} of the vacancy-mediated diffusion on the ladders is lengthy and cumbersome, so the advantage of our continuous approach (in addition to its applicability to arbitrary short-ranged interaction potentials) is even more apparent.

The heuristic argument presented in this section does not depend on whether the interaction potential is attractive or repulsive. Consequently, we expect the results to hold irrespective of the sign of the force on the driven particle. This is shown in Fig.~\ref{fig:ydist_sign} for $d=1$, where we compare $P(y,t)$ measured from numerical simulations for attractive ($V_0<0$) and repulsive ($V_0>0$) Gaussian interaction \eqref{eq:V_gauss}. The identical distribution in both the cases illustrates that the universal scaling function is independent of the sign of the interaction, i.e., whether the interaction is repulsive or attractive.

Although the heuristic approach used in this subsection predicts the exact scaling functions in $d=1$ and $d=2$, the non-universal factors $c_1$ and $c_2$ remain unknown. We now present asymptotically exact computations in a few explicit examples which allow the determination of these non-universal factors.
The emerging scaling functions are the same, supporting the validity of the heuristic argument presented in this subsection.

\subsection{Exact computations: Explicit examples}
\label{sec:exact_dist}

For short-ranged interactions in $d=1$, the Brownian functional form \eqref{eq:yt_approx} allows for the exact computation of the position distribution $P(y,t)$,  using the Feynman-Kac formalism. In this subsection, we derive the scaling function for a few simple interaction potentials.

To compute the full distribution $P(y,t)$ for $y$ at fixed time $t$, we consider the `Laplace transform' of  $P(y,t|x_0)$, the distribution of $y$
at time $t$, given that $x(0)=x_0$, with respect to $y$
\begin{eqnarray}
Q_p(x_0,t) &=& \Big\langle e^{-p \int_0^t f(x(\tau)) d\tau} \Big\rangle \nonumber \\
&=& \int_{-\infty}^{\infty} dy \,e^{-p y} P(y,t|x_0),
\label{eq:Qp}
\end{eqnarray}
where $p$ denotes the conjugate variable.  Note that the integration is over $-\infty<y<+\infty$, as opposed to $0$ to $\infty$ used for the standard Laplace transform. One can show \cite{BF_2005} that $Q_p(x_0,t)$ in Eq.~\eqref{eq:Qp} satisfies the backward Feynman-Kac equation (where the starting point $x_0$ is treated as a variable):
\bea
\partial_t Q_p(x_0,t)= D \partial_{x_0}^2 Q_p(x_0,t) - p f(x_0) Q_p(x_0,t). \label{eq:FK}
\eea
The function $Q_p(x_0,t)$ must satisfy the initial condition $Q_p(x_0,0)= 1$. This assertion follows from the definition in \eqref{eq:Qp} and
the boundary conditions
\bea
Q_p(x_0\to \pm \infty, t)=1.
\label{eq:Q_bc}
\eea
Indeed, if the particle starts at $\pm \infty$, at any finite time $t$, the value of $f(x)$ in Eq.~\eqref{eq:Qp} is essentially zero since we assume that $f(x)\to 0$ as $x \to \pm \infty$. To find $P(y,t)$, we first need to solve Eq.~\eqref{eq:FK} for arbitrary $x_0$ and then set $x_0=0$ and invert the Laplace transform in Eq.~\eqref{eq:Qp}.

To solve Eq.~\eqref{eq:FK}, it is appropriate to take first the Laplace transform with respect to $t$ by defining
\bea
\tilde{Q}_p(x_0,s)= \int_0^\infty dt \, e^{-s t} Q_p(x_0,t)    \label{eq:Qps}
\eea
Taking the Laplace transform of Eq.~\eqref{eq:FK} and using the initial condition $Q_p(x_0,0)=1$ gives
\begin{subequations}
\label{eq:Qps_diff}
\begin{align}
\label{LQps}
&\mathcal{L} \tilde{Q}_{p}(x_0,s)=-1\\
\label{L:def}
&\mathcal{L} = D \partial_{x_0}^2 - s - p f(x_0)
\end{align}
\end{subequations}
The boundary conditions 
\bea
\tilde{Q}_p(x_0\to \pm \infty, s)= \frac 1s \label{eq:Qps_bc}
\eea
follow from \eqref{eq:Q_bc}. In general, Eq.~\eqref{eq:Qps_diff} is hard to solve for arbitrary $f(x_0)$. So, we consider  a couple of specific examples where $\tilde Q_p(x_0,s)$ can be computed explicitly.

\subsubsection{Square potential}
\label{sec:squareV}

For the square potential
\bea
V(x)=\left \{
\begin{split}
V_0 &  \quad \text{for} \quad  -R\le x\le R \cr
    0   & \quad \text{for} \quad  |x|>R    
\end{split}   
\right .               \label{eq:Vx}
\eea
the force is $f(x) = -V_0 \delta(x+R) + V_0 \delta(x-R)$. We thus need to solve Eq.~\eqref{LQps} with 
\begin{align}
\mathcal{L} = D \partial_{x_0}^2 - s-p V_0[\delta(x_0-R)- \delta(x_0+R)].
\label{eq:Qps_box}
\end{align}
The solution is presented in Appendix \ref{app:square_pot}. In particular,  defining
\begin{equation}
\mathcal{Q}_p(s)= \tilde{Q}_p(x_0=0,s)\, ,
\label{def_Qp.1}
\end{equation}
we find the exact expression 
\begin{align}
\mathcal{Q}_p(s)= \frac 1s  \left[ 1- \frac{2\, p^2\, V_0^2\, \sinh(R\, \sqrt{s/D}\,)}{p^2\, V_0^2\,\sinh(2R \sqrt{s/D}\,)- 
2sD\, e^{2 R\sqrt{s/D}}  } \right]\, .
\label{eq:Qp0s_sol}
\end{align}
Expanding $\mathcal{Q}_p(s)$ in powers of $p$ yields
\begin{subequations}
\begin{align}
\label{moment_exp}
&\mathcal{Q}_p(s)= \frac 1s +\sum_{n\geq 1}\rho_n(s)\, p^{2n}\\
&\rho_n(s) =  \frac{V_0^{2n}}{s(sD)^{n}}\,
\left( \frac{\sinh(2 R \sqrt{s/D})}{2}\right)^{n-1} \cr
& \qquad \qquad \times \sinh \left (R\sqrt{\frac sD} \right)\, e^{-2 n R \sqrt{s/D}},
\end{align}
\end{subequations}
where $(2n)!\, \rho_n(s)$ represents the Laplace transform of the $(2n)$-th moment of $y(t)$. All odd moments vanish due to symmetry.
The coefficient $\rho_1(s)$ of the first term of the expansion in $p^2$ has the small $s$ behavior 
\begin{equation}
\rho_1(s)=\frac{V_0^2}{s^2 D}\,\,\sinh(R \sqrt{s/D})e^{-2 R \sqrt{s/D}}\simeq \frac{V_0^2 R}{(sD)^{3/2}}
\end{equation}
from which we deduce (performing the inverse Laplace transform) the large-time asymptotic of the MSD of the passive particle driven by the square potential
\bea
\la y^2(t) \ra \simeq \frac{4 R V_0^2}{\sqrt{\pi D^3}} \sqrt{t}\,. \label{eq:y2av_square}
\eea

To derive the full distribution $P(y,t)$ in the large time regime, we consider the limit $s\to 0$ and $p\to 0$ in \eqref{eq:Qp0s_sol}. To the leading order
we get,
\bea
\mathcal{Q}_p(s) \simeq \frac{1}{\sqrt{s}( \sqrt{s}- c_1^2 p^2)}. \label{eq:Qpx0s}
\eea
where $c_1^2 = V_0^2 R/D^{3/2}$. The large-time position distribution is given by the double inverse Laplace transform of $\mathcal{Q}_p(s)$. This can be done using a set of integral representations [see Appendix~\ref{app:square_pot} for the details]. First, we invert the Laplace transform with respect to $p$ and find,
\begin{align}
\hat P(y,s) \simeq \int_0^\infty \frac{dN}{\sqrt{4 \pi c_1^2 Ns}}\,\,\exp{\Big[- N\sqrt{s} -\frac{y^2}{4 c_1^2 N}\Big]}. \label{eq:P_ys}
\end{align}
Then, we invert the Laplace transform with respect to $s$ to get,
\begin{align}
P(y,t) \simeq \int_0^{\infty} \frac{dN}{\sqrt{4 \pi^2 c_1^2 Nt}}\,\,\exp{\Big[-\frac{y^2}{4c_1^2N}- \frac{N^2}{4t}\Big]}. \label{eq:Pyt_int}
\end{align}
Making a change of variable $N= u\sqrt{4t}$, we get the scaling form for the position distribution
\bea
P(y,t) \simeq \frac 1{c_1 t^{1/4}} {\cal F}_1 \Big(\frac y{c_1 t^{1/4}} \Big).  \label{eq:Pyt_scaling}
\eea
The scaling function ${\cal F}_1(z)$ is the same as predicted by the heuristic argument, viz. it is given by the integral in Eq.~\eqref{eq:1d_int_SF} that reduces to \eqref{eq:1d_SF}. The exact solution also provides the non-universal scale factor $c_1= V_0\sqrt{R}/D^{3/4}$.

\subsubsection{Tent potential}
\label{sec:linV}

For the tent potential \eqref{eq:V_lin}, the corresponding force in $d=1$ is
\bea
f(x) =  \left \{ 
\begin{split}
\text{sgn}(x) \frac{V_0}{R} & \quad \text{for}~~ |x| < R \cr
 0 &  \quad \text{otherwise}. 
   \end{split}
 \right. \label{eq:fx_linear}
\eea
In this case, the backward Fokker-Planck equation is also solvable, see Appendix \ref{app:Linear_potential}. As before, we compute the second moment by expanding $\mathcal{Q}_p(s)$ around $p=0$, and inverting the Laplace transform of the coefficient of $p^2$, we obtain
\bea
 \la y^2(t) \ra \simeq  \frac {4RV_0^2}{3 \sqrt{\pi D^3}}\, \sqrt{t}\,.
 \label{eq:y2t_lin}
\eea
To derive the position distribution $P(y,t)$ of the driven particle we use the same procedure as before [see Appendix \ref{app:Linear_potential} for details] and find 
\bea
\mathcal{Q}_p(s) \simeq \frac 1 {\sqrt{s} (\sqrt{s}- c_1^2 p^2)} \label{eq:Qps_lin}
\eea
with $c_1^2=R V_0^2/(3D^{3/2})$ when  $s\to 0$ and $p\to 0$. This formula differs from Eq.~\eqref{eq:Qpx0s} only by the nonuniversal numerical factor $c_1$. Thus, as expected, the scaling form \eqref{eq:Pyt_scaling} is the same for both potentials.

\begin{figure}[t]
\includegraphics[width=7.5 cm]{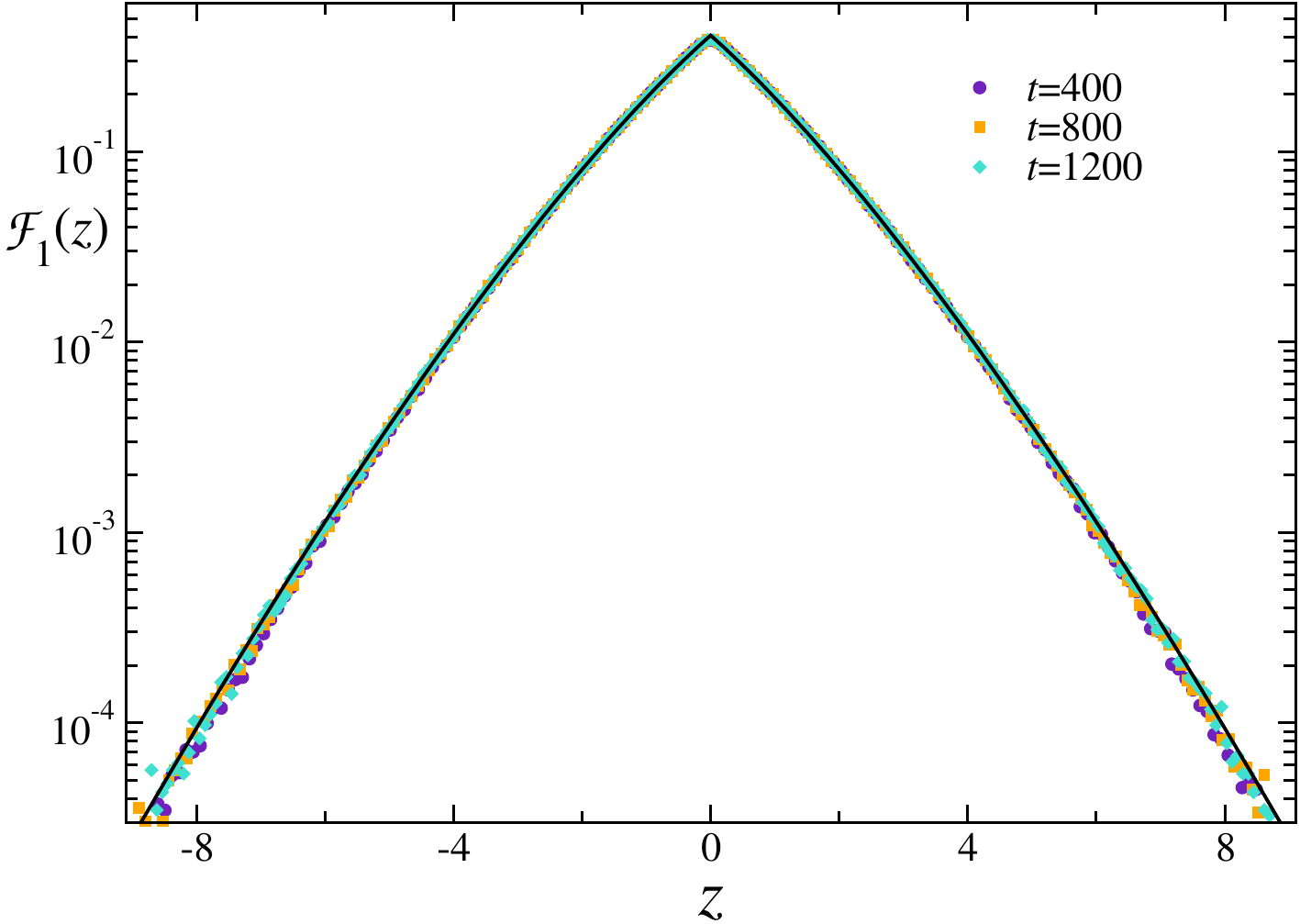}
\caption{Distribution of the scaled variable $z=y/(c_1t^{1/4})$ for the tent potential \eqref{eq:V_lin} with $c_1 = V_0 \sqrt{R}/D^{3/4}$. The symbols denote the data obtained from numerical simulations with $R=0.5$ and $V_0 =1.0$, and the solid line indicates the scaling function \eqref{eq:1d_SF}. Here we have taken $D=1$.} 
\label{fig:1d_dist_Vlin}
\end{figure}

Figure~\ref{fig:1d_dist_Vlin} shows a comparison of the scaled position distribution measured from numerical 
simulations of the passive particle driven by the tent potential with the theoretical prediction 
\eqref{eq:1d_SF} for the scaling function.

\section{Long-Ranged interaction}
\label{sec:long_range}

In this section, we focus on long-ranged interaction between the driver and the driven particle, as in \eqref{FA}. For simplicity, we limit ourselves here to $d=1$, and consider long-ranged forces that decay as
\begin{equation}
\label{long-ranged.1}
f(z) \simeq \text{sgn}(z)\frac{A}{|z|^a} \quad\text{as}\quad |z|\to\infty,
\end{equation}
with the exponent $a>0$. The force can be either repulsive $(A>0)$ or attractive $(A<0)$.
Below we focus only on the MSD of the driven particle and show that at late times
it grows algebraically as $\sim t^{\phi}$ where the exponent $\phi$ 
depends on the range of $a$, as announced in \eqref{phi}. Hence, we consider
the different ranges of $a$ separately below.

\subsubsection{$a>1$: sub-diffusive growth.}

To see whether the MSD of the driven particle shows any universal behavior for long-ranged interactions, we start with \eref{eq:y2_laplace}, which gives the Laplace transform of the MSD. We can rely on \eref{eq:y2_laplace} only when the typical displacement of the Brownian particle greatly exceeds a typical 
displacement of the driven particle, $\langle y^2(t)\rangle \ll t$. In other words, the driven particle exhibits sub-diffusive growth at late times. 
We will justify this condition {\it a posteriori}.

We employ the same arguments as in the short-ranged interaction case. To extract from \eref{eq:y2_laplace} 
the small $s$ behavior of the Laplace transform $M_2(s)$ of the MSD of the driven particle, we need the 
small $k$ behavior of the Fourier transform of the force, $\tilde f(k)$. For long-ranged interaction of the type in \eqref{long-ranged.1}, the
Fourier transform $\tilde f(k)$ is non-analytic around $k=0$ for $a<2$ and its leading order behavior is 
\bea 
\tilde f(k) \sim 
\begin{cases}
k & \text{for}~~ a>2 \cr
|k|^{a-1} & \text{for}~~ a \le 2.
\end{cases}
\label{eq:lr_fk}
\eea 
In this case, the small $s$ behavior of $M_2(s)$ is given by [see Appendix \ref{ap:msd} for the details]
\bea
M_2(s) \sim 
\begin{cases}
s^{a -3} & \text{for} \quad 1\le a \le \frac{3}{2} \cr
s^{-3/2} & \text{for} \quad a > \frac 32.
\end{cases} \label{eq:msd_s_long}
\eea
Correspondingly, the long-time asymptotic behavior of the MSD of the driven particle is given by
\bea
\la y^2(t) \ra \sim  t^\phi \quad {\textrm as}\quad t\to \infty\, ,
\label{MSD.1}
\eea 
with the exponent
\bea
\phi =  \max\big(2-a,\tfrac{1}{2}\big).
\label{eq:msd_t_long}
\eea
The exponent $\phi$ characterizing the growth of the MSD of the driven particle undergoes a `freezing 
transition'--  $\phi=2-a$ is a strictly decreasing function of $a$ in the range 
$1<a<\frac{3}{2}$, and it freezes at the value $\phi=\frac{1}{2}$, the same as for 
short-ranged interactions, when $a$ crosses the critical value $a_c=\frac{3}{2}$. The motion of 
the driven particle is thus sub-diffusive, $\phi<1$, for any $a>1$, so our assumption $\langle y^2(t)\rangle \ll t$ is valid in the long time limit justifying the usage of \eref{eq:y2_laplace}.
This proves the result announced in the first two lines of \eqref{phi}.

\begin{figure}
\includegraphics[width=7.0 cm]{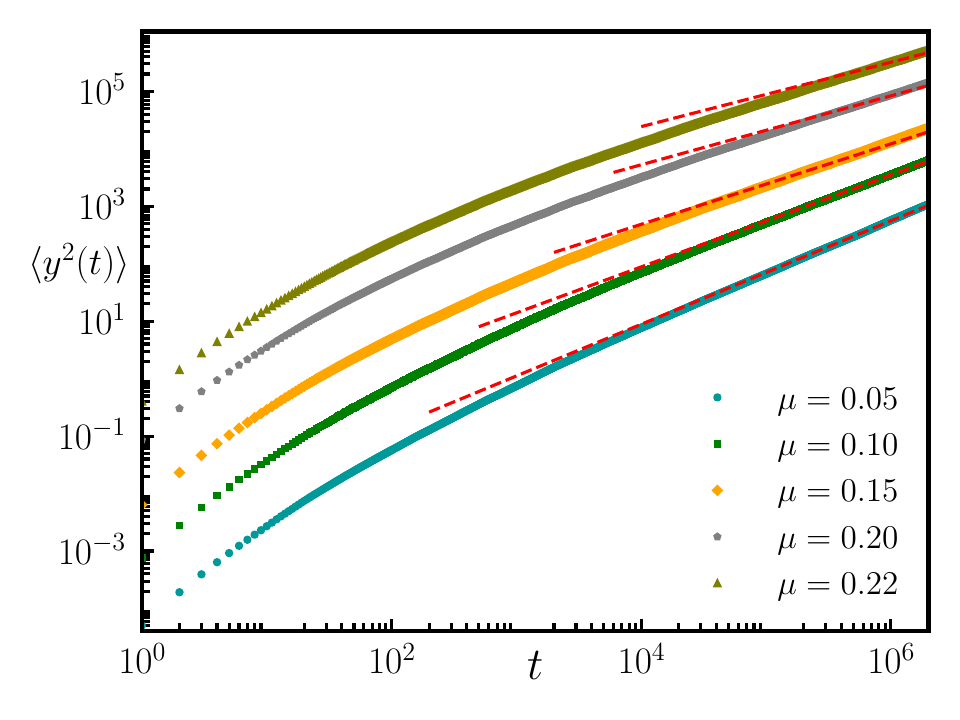}
\caption{Long-ranged interactions. The plot of the MSD of the passive particle driven by the power-law interaction potential \eqref{eq:V_long}. The symbols denote the data from numerical simulations, and the red dashed line indicates the predicted asymptotic behavior \eqref{eq:msd_t_long} for $\mu< 1/4$. Here we have taken $V_0=1=D$.} \label{fig:long_msd}
\end{figure}

We now illustrate this general result for $a>1$ by focusing on a class of 
interaction potentials
\bea
V(z) = \frac{V_0}{(1+z^2)^\mu}\,,
\label{eq:V_long}
\eea
parametrized by the exponent $\mu>0$.
The force associated to this potential is long-ranged, 
\begin{equation}
f(z) =\frac{2 \mu V_0 z}{(1+z^2)^{(1+\mu)}}\, ,
\label{ex.1}
\end{equation}
which decays for large $|z|$ as $|z|^{-(1+2\mu)}$. Hence, it corresponds to an example
of a force of the type in \eqref{long-ranged.1} with exponent
\begin{equation}
a= 1+ 2\mu\, .
\label{amu.1}
\end{equation}
Since $\mu>0$, this corresponds to $a>1$ where we expect the result for MSD in \eqref{MSD.1} and
\eqref{eq:msd_t_long} to hold.
It turns out that the Fourier transform of the force in \eqref{ex.1} can be written explicitly as
\bea
\tilde f(k) = \frac{i\, V_0\, 2^{\frac 32 - \mu}\, \sqrt{\pi} }{\Gamma(\mu)}\, \text{Sgn}(k)\, 
|k|^{\mu + \frac 12}\, K_{\mu - \frac 12}(|k|)\, , 
\label{eq:fk_long}
\eea
where $K_\nu(z)$ denotes the modified Bessel function of the second kind of order $\nu$.

\begin{figure}
\includegraphics[width=8.8 cm]{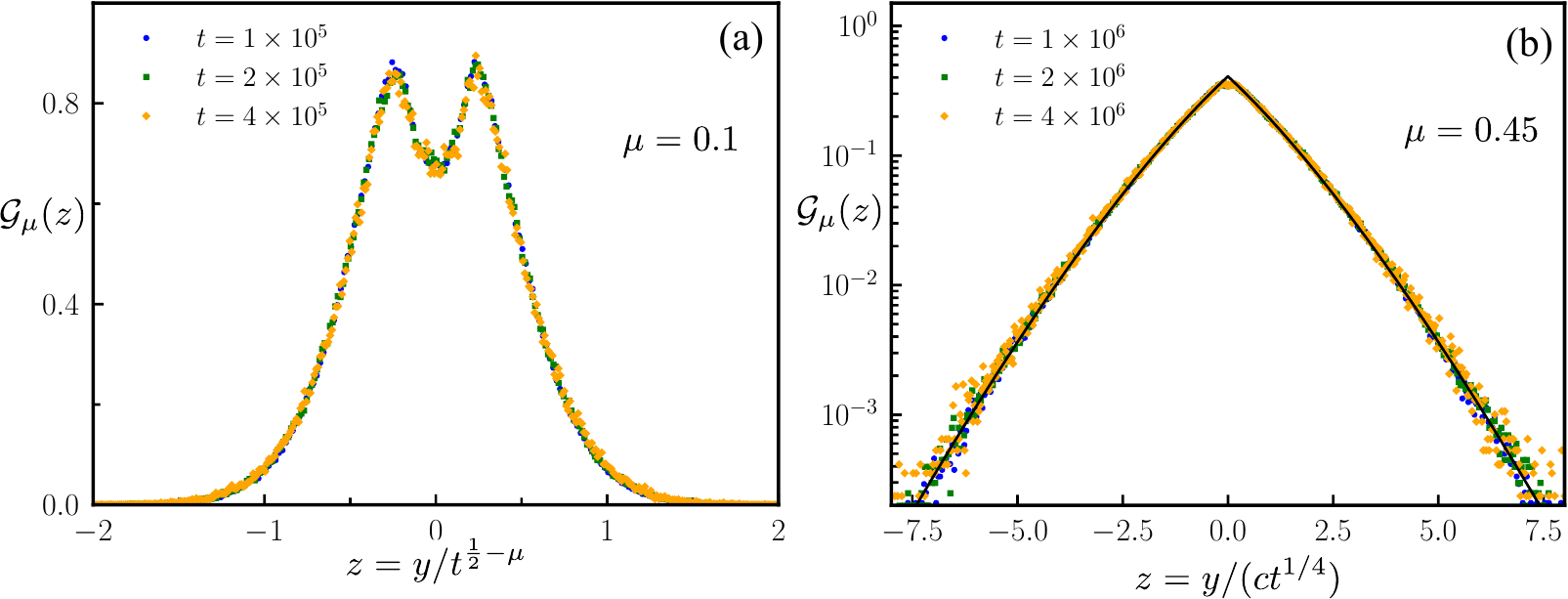}
\caption{Long-ranged interactions: Data collapse of the position distribution of the passive particle driven by the power-law interaction potential \eqref{eq:V_long} obtained from numerical simulations for (a) $\mu<1/4$ and (b) $\mu > 1/4$. (a) Distribution of the scaled position $z= y t^{\mu - \frac 12}$  obtained from numerical simulation with $\mu=0.1$. (b) Distribution of the scaled position $z= y/(ct^{1/4})$, with an arbitrary scaling factor $c$, for $\mu=0.45$. The solid line shows the scaling function \eqref{eq:1d_SF}. Here we have taken $V_0=1$ and $D=1$.} \label{fig:dist_long}
\end{figure}

As discussed before, it suffices to look at the behavior of $\tilde f(k)$ for small $k >0$. To extract this 
asymptotic, we need the small $k$ behavior of the modified Bessel function of the second kind \cite{Bender}
\begin{align}
\label{Knu}
K_{\nu}(k) \simeq \frac 12 \left[\left(\frac k2 \right)^{-\nu} \Gamma(\nu) + \left(\frac k2 \right)^{\nu} \Gamma (-\nu)\right].
\end{align}
This equation implies that the leading small $k$ behavior of the modified Bessel function $K_{\mu - \frac 
12}(k)$ is different depending on whether $\mu > 1/2$ or not. Substituting the asymptotic \eqref{Knu} into 
\eref{eq:fk_long} we obtain the anticipated algebraic behavior, 
\bea
\label{alpha-mu}
\tilde f(k) \sim 
\begin{cases}
|k|^ {2\mu} & \text{for}~~ \mu < \frac 12 \cr
k        & \text{for}~~ \mu > \frac 12
\end{cases}
\eea
which is consistent with Eq.~\eqref{eq:lr_fk} with $a=1+2\mu$ [see Eq.~\eqref{amu.1}]. Consequently, the MSD of the passive particle 
driven by the long-ranged potential \eqref{eq:V_long} scales as
\bea
\frac{\la y^2(t) \ra}{V_0^2} \sim 
\begin{cases}
t^{1-2 \mu} & \text{for}~~ \mu < \frac 14 \cr
\sqrt{t} & \text{for}~~ \mu > \frac 14
\end{cases} 
\label{eq:phi_mu}
\eea
in the long time limit, with the `freezing transition' occurring at $\mu=\frac{1}{4}$. Using $a=1+2\mu$, this translates precisely into the result in \eqref{eq:msd_t_long}. This behavior is illustrated in Fig.~\ref{fig:long_msd} where this analytical prediction is compared with the data obtained from numerical simulations, which shows an excellent agreement.

To explore the signature of the freezing transition on the full position distribution, we studied the position distribution $P(y,t)$ numerically. The data support the emergence of the scaling in the long time 
limit:
\bea 
P(y,t) \simeq \frac 1{\sqrt{\la y^2(t) \ra}} \, {\cal G}_\mu \left (\frac y {\sqrt{\la y^2(t) \ra}} \right). 
\label{eq:G_mu}
\eea   
The freezing transition at $\mu=\frac{1}{4}$ is associated with a shape transition of the scaling function. 
For $\mu > 1/4$, as already indicated by Eq.~\eqref{eq:phi_mu}, the behavior of the driven particle is 
similar to that with a short-ranged interaction and hence ${\cal G}_\mu(z) = {\cal F}_1(z)$, i.e., the 
scaling function is given by Eq.~\eqref{eq:1d_SF}. For $\mu < \frac{1}{4}$, the distribution depends 
continuously on $\mu$ and has a bimodal shape; the scaling function in this regime is unknown. We illustrate the shape transition in Fig.~\ref{fig:dist_long}, where we plot the scaled position distribution for two 
different values of $\mu$ in the two regimes.

So far, we have tacitly assumed that $\mu\ne \frac{1}{4}$ and $\mu>0$. We now analyze the behavior in the marginal 
cases $\mu=\frac{1}{4}$ and $\mu=0$, and also the behavior when $\mu<0$. This range may seem unphysical, but 
recalling that the force scales as $z^{-(1+2\mu)}$ when $z\to\infty$ we conclude that the force 
asymptotically vanishes when $\mu>-\frac{1}{2}$. Below, we use the potential \eqref{eq:V_long} characterized 
by the exponent $\mu$ and a much more general class of potentials with force decaying as $z^{-a}$. The 
qualitative results are identical if we choose $a=1+2\mu$. In some calculations, we need the entire 
potential, not only the tail, and we use the potential \eqref{eq:V_long} in these situations.


\vskip 0.2cm

\begin{figure}
\includegraphics[width=6 cm]{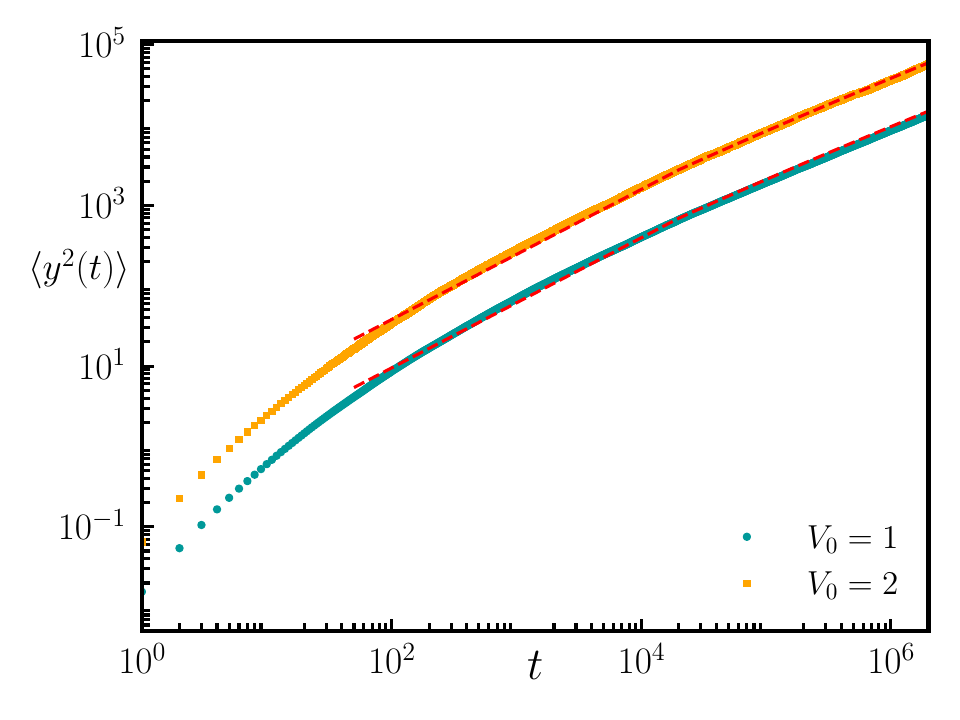}
\caption{Long-ranged interaction with $a=3/2$: Plot of the MSD of the passive particle driven by the power-law interaction potential \eqref{eq:V_long} for the marginal case $\mu=1/4$ (i.e., $a=3/2$), for different values of $V_0$. The symbols denote the data from numerical simulations, and the red dashed lines indicate the predicted asymptotic behavior \eqref{MSD:marginal}. Here we have taken $D=1$.} \label{fig:msd_long_a1.5}
\end{figure}

\noindent {\bf {The marginal case $a=3/2$:}}
For $\mu=\frac{1}{4}$ or equivalently $a=3/2$ in \eqref{amu.1}, the leading small $k$ behavior of the 
modified Bessel function $K_0(k)$ reads
\begin{align}
\label{K0}
K_0(k) \simeq \log\tfrac{2}{k} - \gamma_E
\end{align}
and Eq.~\eqref{eq:fk_long} becomes 
\bea
\tilde f(k) \simeq 2\, i\, V_0\,  k\, \big[\log\tfrac{2}{k} - \gamma_E\big]. 
\label{eq:fk_short}
\eea
The same analysis as before [see Appendix~\ref{ap:msd}] gives the small $s$ behavior $M_2\sim s^{-3/2}(\log 
s)^2$ of the Laplace transform of the MSD of the driven particle, from which the large-time behavior is
\begin{equation}
\label{MSD:marginal}
\la y^2(t) \ra \sim V_0^2\, \sqrt{t}\,(\log t)^2\,.
\end{equation}
In addition to the announced temporal scaling \eqref{y2:log}, we included the dependence on $V_0$ but omitted an exact amplitude as we omitted an exact $\mu-$dependent amplitude in Eq.~\eqref{eq:phi_mu}. The long-time behaviour of MSD for the $a=3/2$ case is illustrated in Fig.~\ref{fig:msd_long_a1.5}.

Thus summarizing, for the passive particle driven by a 
long-ranged force decaying as $z^{-a}$, we can restate growth laws \eqref{eq:phi_mu} and 
\eqref{MSD:marginal} as
\begin{equation}
\label{MSD:long}
\la y^2(t) \ra \sim 
\begin{cases}
\sqrt{t}                      &  a>\frac{3}{2} \\
\sqrt{t}\,(\log t)^2      &  a=\frac{3}{2} \\
t^{2-a}                      &  1<a<\frac{3}{2}
\end{cases}
\end{equation}
The exponent $\phi$ cited in \eqref{phi} for $a>1$ immediately follows from \eqref{MSD:long}. 

\subsubsection{$0<a<1$}

To complete the growth laws of the MSD, we now probe the behavior in the $0<a\leq 1$ range. Unfortunately, we can no longer rely on the methods employed so far. Indeed, the driven particle is expected to be as mobile as the Brownian particle, or even super-diffusive, so the chief starting point of our analysis, the Brownian functional \eqref{eq:yt_approx}, is no longer applicable. We can advance, however, once we recognize the crucial 
distinction between the attractive $A<0$ and the repulsive $(A>0)$ long-ranged forces in \eqref{long-ranged.1} with exponent $a$ in the range $0<a<1$.

If the force is repulsive $(A>0)$, the passive particle eventually settles on one of the two half-lines, $y>0$ or 
$y<0$, and remains there. Both outcomes are equiprobable if the starting positions coincide. In the former 
case, we anticipate that the passive particle is pushed to a distance $y\gg \sqrt{Dt}$. In this case, the force $A/z^a$ [cf. 
Eq.~\eqref{long-ranged.1}] simplifies to $A/y^a$. Therefore, the average position satisfies the deterministic equation
\begin{equation}
\label{FA:det}
\dot y = \frac{A}{y^a}
\end{equation}
in the leading order, from which
\begin{equation}
\label{repulsive}
y = [A\,(a+1)\, t]^\frac{1}{a+1}
\end{equation}
where $A>0$ ince the force is repulsive) and $0<a<1$. This provides the third line of the result announced in \eqref{phi}. Thus in this case, the motion of the driven particle is asymptotically deterministic, albeit the direction of the motion is random. The scaling exponent obeys 
$\frac{1}{2}<\frac{1}{a+1}<1$ when $0<a<1$, so the motion is super-diffusive and sub-ballistic. Let us now specialize to the potential in \eqref{eq:V_long} or equivalently to the force
in \eqref{ex.1} which, for large $z$, decays as
$f(z)\simeq 2 \mu V_0\, z^{-(1+2\mu)}$, indicating that $A=2 \mu V_0$ and $a=1+2\mu$ in \eqref{FA:det}.
Since $0<a<1$, this indicates that $-1/2<\mu<0$.
In order that the force is repulsive, i.e., $A>0$, we must have $2\, \mu, V_0>0$. Since
$-1/2<\mu<0$ is negative, we must choose $V_0<0$ to have a repulsive force.
The result in \eqref{repulsive} then states
\begin{equation}
\label{repulsive:V}
y = [4\,V_0 \, \mu\, (\mu+1) t]^\frac{1}{2\,(\mu+1)}\, ,
\end{equation}
applicable when $0>\mu>-\frac{1}{2}$ and $V_0<0$.

Different behaviors of the driven particle emerge when the force is attractive. Subtracting \eqref{passive} 
from \eqref{active} we obtain a stochastic differential equation
\begin{equation}
\dot z = f(z) - \eta,
\end{equation}
for the separation $z=y-x$. The probability distribution $P(z,t)$ satisfies the Fokker-Planck equation 
\begin{equation}
\label{FPE}
\frac{\partial P(z,t)}{\partial t} = - \frac{\partial }{\partial z}\left[f(z) P(z,t)\right] + 
D\,\frac{\partial^2 P(z,t)}{\partial z^2}\,.
\end{equation}
The crucial observation is that for slowly decaying attractive forces \eqref{FA} with exponent satisfying 
$0<a<1$, the driven particle is `slaved' to the Brownian particle which is mathematically reflected by the 
fact that the solution of the Fokker-Planck equation becomes stationary in the long-time limit. Seeking the 
stationary solution we reduce \eqref{FPE} to
\begin{equation}
\frac{d P(z) }{d z} = \frac{f(z)}{D}\,P(z) 
\end{equation}
which is solved to yield
\begin{equation}
\label{Pz:sol}
P(z) = C\,\exp\!\left[\int_0^z dw\,\frac{f(w)}{D}\right]
\end{equation}
with amplitude $C$ fixed by normalization. Specializing \eqref{Pz:sol} to the potential \eqref{eq:V_long} we 
obtain
\begin{equation}
\label{Pz:sol-V}
P(z) = C\,\exp\!\left[-\frac{V_0}{D(1+z^2)^\mu}\right]
\end{equation}
applicable when $0>\mu>-\frac{1}{2}$ and $V_0>0$ (since the force is attractive). 

Since the driven particle is slaved to the Brownian particle, the MSD of the driven particle is
\begin{equation}
\label{MSD:attractive}
\la y^2(t) \ra \simeq 2Dt 
\end{equation}
in the leading order. The asymptotic behaviors \eqref{repulsive} and \eqref{MSD:attractive} imply the 
predictions for the exponent $\phi$ in the $0<a<1$ range announced in Eq.~\eqref{phi}.

\subsubsection{$a=1$}

Finally, we discuss the marginal case $a=1$. The entire force distribution is necessary if the force is 
attractive [cf. Eq.~\eqref{Pz:sol}], and as an illustration we take
\begin{equation}
\label{F-log}
f(z) = - \frac{2\,A\,D\,z}{1+z^2}\,.
\end{equation}
The corresponding potential 
\begin{equation}
\label{Coulomb}
V(z) = A\,D\,\ln(1+z^2)
\end{equation}
is a regularized $\mu=0$ version of the potential \eqref{eq:V_long}. Alternatively, this potential is a 
regularized two-dimensional Coulomb potential, although we use it in one dimension. We put the diffusion coefficient $D$ 
into \eqref{F-log}, so the amplitude $A$ is dimensionless. The force \eqref{F-log} is used in numerous 
applications, particularly in modeling the spreading of the momentum of two-level atoms in optical lattices 
\cite{Castin,Zoller,Lutz} where $z$ is the momentum of an atom, $D$ describes momentum fluctuations which 
lead to heating, and $f(z)$ the cooling force. Hence, the Fokker-Planck equation with force \eqref{F-log} was the subject of several articles 
\cite{bray2000random,Castin,Zoller,Lutz,Chavanis,levine2005long,Douglas,KB10,
hirschberg2011approach,dechant2011solution,ray2020diffusion}; see Ref.~\cite{Barkai} for a review. In applications to laser cooling, $A>0$, so in our terminology, the force is attractive. Hence, we can use known results in this case. A discrete lattice model of a particle moving in a logarithmic
potential was introduced by Gillis in the probability 
literature \cite{gillis1956centrally}. For more recent studies
on the Gillis model and its generalizations, 
see Refs.~\cite{onofri2020exploring, pozzoli2020continuous, radice2020statistics, artuso2022extreme,zodage2023sluggish}.
Here we focus on the continuous-space version of the model,
briefly re-derive a few simple results using our terminology and 
notations and cite more complicated results.

The force \eqref{F-log} is attractive when $A>0$, and the stationary solution becomes
\begin{equation}
P(z) = C (1+z^2)^{-A}
\end{equation}
This solution is normalizable when $A>\frac{1}{2}$. In this case we fix the amplitude from the normalization 
requirement, $\int_{-\infty}^\infty dz\,P(z)=1$, and arrive at
\begin{equation}
\label{Pz}
P(z) = \frac{\Gamma(A)}{\Gamma\big(\frac{1}{2}\big)\,\Gamma\big(A-\frac{1}{2}\big)} \,(1+z^2)^{-A}
\end{equation}
The precise form \eqref{Pz} depends on the detailed form \eqref{F-log} of the force, but the threshold value 
$A_*=\frac{1}{2}$ is universal. This assertion readily follows from the general stationary solution 
\eqref{Pz:sol}.

The stationary solution \eqref{Pz} was noticed in \cite{Lutz,Douglas}, and it was observed in optical 
lattice experiments \cite{Douglas}. Notice, however, that the stationary solution \eqref{Pz} predicts finite 
second moment $\langle z^2\rangle$ only when $A>\frac{3}{2}$. The divergence of the second moment was 
noticed in \cite{KB10} where it was shown that when $\frac{1}{2}<A\leq \frac{3}{2}$, the stationary solution 
\eqref{Pz} is valid only in the central region $|z|\ll \sqrt{t}$; in the tail regions the solution is 
time-dependent. The behavior of the second moment can be deduced from a more comprehensive analysis 
\cite{KB10} of the Fokker-Planck with force \eqref{F-log}:
\begin{equation}
\label{z2:A}
\langle z^2\rangle = 
\begin{cases}
(2A-3)^{-1}                      & A>\frac{3}{2}\\
\sim t^{\frac{3}{2}-A}       & \frac{1}{2}<A < \frac{3}{2}\\
2D(1-2A)t                        & 0<A<\frac{1}{2}        
\end{cases}
\end{equation}
We observe that when the force \eqref{F-log} is weakly attractive, $0<A<\frac{1}{2}$, the separation between 
the driver and the passive particle grows diffusively. This suggests to seek a scaling solution of the 
Fokker-Planck equation with force \eqref{F-log}:
\begin{subequations}
\begin{equation}
\label{Pzt:scal}
P(z,t)=\frac{1}{\sqrt{4Dt}}\,\mathcal{P}(\zeta), \qquad \zeta = \frac{z}{\sqrt{4Dt}}\,. 
\end{equation}
In the range $0<A<\frac{1}{2}$, the scaled distribution 
\begin{equation}
\label{Pzeta}
\mathcal{P}(\zeta) = \frac{1}{\Gamma\big(\frac{1}{2}-A\big)}\,\zeta^{-2A}\,e^{-\zeta^2} 
\end{equation}
\end{subequations}
was found in Ref.~\cite{KB10}. The same scaling solution applies repulsive forces, $A<0$. 

Using \eqref{Pzt:scal}--\eqref{Pzeta} one finds the diffusive growth 
\begin{equation}
\label{z2}
\langle z^2\rangle = 2\,D\,(1-2A)\,t
\end{equation}
asymptotically valid for $A<\frac{1}{2}$, i.e., for sufficiently weak attractive forces \eqref{F-log} as it 
was already stated in \eref{z2:A} and for all repulsive forces \eqref{F-log}. In this range, the MSD of the 
driven particle is also expected to grow diffusively
\begin{equation}
\label{MSD:A}
\la y^2(t) \ra \simeq C(A)\, D\, t 
\end{equation}
The amplitude $C(A)$ is unknown, apart from the obvious case of vanishing force when $y\equiv 0$ and hence 
$C(0)=0$.  The identity $\la z^2(t) \ra = \la x^2(t) \ra - 2\la x(t)y(t) \ra + \la y^2(t) \ra$ and the 
knowledge of $\la z^2(t) \ra$, Eq.~\eqref{z2}, and $\la x^2(t) \ra = 2Dt$ is insufficient since in this 
marginal case the stochastic variables $x(t)$ and $y(t)$ are correlated and, in particular, $\la x(t)y(t) 
\ra\ne 0$. The correlator cannot vanish because this would lead to $C(A)=-4A$ which is clearly impossible 
for weakly attractive forces: $0<A<\frac{1}{2}$. Only for very strong repulsive forces \eqref{F-log} the 
naive prediction is asymtotically exact: $C(A)\simeq -4A$ when $A\to -\infty$.

\section{Concluding Remarks}
\label{sec:concl}

We have studied the motion of a passive particle driven by a non-reciprocal interaction with a Brownian particle. We find that the late-time position fluctuations of the driven particle are remarkably universal when the interaction potential is short-ranged. The MSD of the driven particle scales as $\sqrt{t}$ in $d=1$ and as $\log t$ in $d=2$, and the position distributions are also universal and independent of the specific form of the potential that only affects the numerical scale factor.  We calculate the scaling functions in $d=1$ and $d=2$ using heuristic arguments. We argue that our calculation is exact, modulo the scale factor, and support our claims by making rigorous calculations for a few specific interaction potentials and by numerical simulations. In $d>2$, the random walk is not recurrent, and therefore, in the case of the short-range interactions, the displacement of the passive particle from its initial position remains
finite in the long time limit. Furthermore, this displacement varies from realization to realization hence, there is no scaling behavior, and the distribution of the final displacement is expected to be non-universal.

We have also explored the motion of a passive particle in one dimension driven by long-ranged interaction potentials. We have found a rich set of behaviors chiefly determined by the exponent $a$ describing the decay of the force, $f\sim z^{-a}$ when $z\to\infty$. The MSD of the driven particle grows as $t^\phi$, with the exponent $\phi$ undergoing a freezing transition at $a=\frac{3}{2}$: The same value $\phi=1/2$ as for short-ranged potentials holds for $a > \frac{3}{2}$, while the exponent varies continuously in the $1<a < \frac{3}{2}$ range where $\phi=2-a$. The behavior remains universal in the $a>1$ range. The scaled position distribution is known only when $a> \frac{3}{2}$, where it coincides with the scaled distribution in a short-ranged potential. The scaled distributions in the $1<a < \frac{3}{2}$ are unknown. The behavior in the $0<a<1$ range is less universal but simpler than when $a\geq 1$. The behavior crucially depends on whether the potential is attractive or repulsive. If the potential is repulsive, the driven particle exhibits asymptotically self-averaging behavior---one of the two possible directions is eventually selected (this varies from realization to realization), and the motion is asymptotically deterministic. If the potential is attractive, the passive particle is effectively slaved to the Brownian particle. We studied the effect of long-ranged potentials in one dimension. The extension to higher dimensions is left for future work.

In this paper, we restricted ourselves to a force $f(r)$ that is regular near $r=0$. It would be interesting to study a singular force that diverges at short distance as a power law $f(r)\simeq B\, r^{-b}$ with a positive exponent $b>0$.

We mentioned the connection between our passive particle driven by a short-ranged non-reciprocal interaction with a Brownian particle and diffusion of a tagged particle mediated by a single vacancy~\cite{Hilhorst,Zoltan,Newman,Gleb02}. The latter problem is traditionally studied in the realm of simple symmetric exclusion process where particles can hop only to nearest neighbors, so in the case of a single vacancy the number of possible moves is $2d$ in the case of the hyper-cubic lattice $\mathbb{Z}^d$. Symmetric exclusion processes in which long-ranged hops are allowed could be a counterpart of our problem with long-ranged potential. In a supplementary direction, a charged tagged particle mediated by a single vacancy exhibits intriguing behaviors \cite{Gleb02} suggesting a similar model in our setting, perhaps just considering a force that is not 
an odd function of the separation.

Arcsine laws are old but still striking characteristics of Brownian motion \cite{levy,IM:book,BM:book}. One can ask about analogues of arcsine laws for the passive particle. For long-ranged forces, the answers could be easier 
than for short-ranged forces. For instance, for forces decreasing as $z^{-a}$ when $z\to\infty$ with exponent 
$0<a<1$, the distribution of the fraction $\tau$ of time spent by the passive particle on the right half-line, $y>0$, is $\frac{1}{2}\delta(\tau)+\frac{1}{2}\delta(\tau-1)$ when the force is repulsive, and the same as for Brownian particle when the force is attractive.

In  this work, we studied the most basic few-body system: A single driver particle and a single passive particle, symbolically a (1,1) system. Similarly, an $(m,n)$ system consists of $m$ Brownian particles and $n$ passive particles supplemented by non-reciprocal forces exerted by every driver particle on every passive particle. The $(\infty, 1)$ system in which non-interacting Brownian particles are uniformly distributed in the $\mathbb{R}^d$ space with  density $\rho$ is an amusing microscopic model of Langevin dynamics. The passive particle should behave 
diffusively. The diffusion coefficient $\mathcal{D}$ of the passive particle depends on the diffusion coefficient 
$D$ of Brownian particles, their density $\rho$, the characteristic scale $V_0$ of the potential, and its range 
$R$. On dimensional grounds, $\mathcal{D}=D\,\Phi(\rho R^d, V_0/D)$, i.e., it depends on two dimensionless 
parameters, the typical number $\rho R^d$ of Brownian particles exerting the force on the passive particle, and $V_0/D$ playing the role of the P\'{e}clet number. Computing $\Phi(\rho R^d, V_0/D)$, e.g., for the Gaussian potential, is a promising avenue for future work.

\acknowledgements 
PLK is grateful to Y. Kafri for a useful discussion. UB acknowledges support from Science and Engineering Research Board (SERB), India under the MATRICS scheme (No. MTR/2023/000392). SNM acknowledges supports from ANR Grant No.~ANR-23-CE30-0020-01 EDIPS and the Science and Engineering Research Board (SERB, Government of India), under the VAJRA faculty scheme (No.~VJR/2017/000110) during a visit to Raman Research Institute, where part of this work was carried out.

\appendix

\section{Large-time behavior of the MSD}
\label{ap:msd}

In this Appendix we provide the detailed derivation of the long-time behavior of the mean-squared displacement. We start with the $d$-dimensional 
version of \eqref{eq:yt_approx}:
\bea
\boldsymbol y(t)\simeq \int_0^t d\tau \,  \boldsymbol f(x(\tau)),  \label{eq:yt_approx_d}
\eea
where $\boldsymbol{f}(\boldsymbol z) = - \nabla V(\boldsymbol z) $ denotes the vector force derived from the central potential $V(|z|)$.  The corresponding mean-squared displacement is 
\bea
\la y^2 (t) \ra = 2 \int_0^t dt_2 \int_0^{t_2} d t_1 \left\langle \boldsymbol{f}(\boldsymbol{x}(t_2)) \cdot \boldsymbol{f}(\boldsymbol{x}(t_1)) \right \rangle
\label{eq:y2_def_d}
\eea
where we have used the fact that the two-point correlation is symmetric under the exchange $t_1  $ and $t_2$.  Hence,  it suffices to compute the two-point correlation for $t_2 > t_1$, which is given by
\begin{align}
 \Big \la \boldsymbol{f}(\boldsymbol{x}(t_2)) \cdot \boldsymbol{f}(\boldsymbol{x}(t_1)) \Big \ra =&  \int d \boldsymbol{x}_2 \int  d \boldsymbol{x}_1 \, f(\boldsymbol{x}_2) f( \boldsymbol{x}_1) \cr 
& \times \mathcal {P}(\boldsymbol{2}| \boldsymbol{1}) \mathcal {P}(\boldsymbol{1}| \boldsymbol{0}) 
\label{eq:fcor_def_d} 
\end{align} 
where {\bf 2}=$(\boldsymbol{x}_2,t_2)$,   {\bf 1}=$(\boldsymbol{x}_1,t_1)$,  {\bf 0}=$(\boldsymbol{0},0)$, and the propagator is given by
\begin{align}
\label{P21}
\mathcal {P}(\boldsymbol{2}| \boldsymbol{1}) = \frac 1{[4 \pi D (t_2-t_1)]^{\frac d2}}\exp{\left[- \frac {(\boldsymbol{x}_2-\boldsymbol{x}_1)^2}{4 D (t_1-t_2)}\right]}
\end{align}
in $d$ dimensions where we have taken $t_2>t_1$.  To compute the two-point correlation, we express the propagator as a Fourier integral and use it in \eref{eq:fcor_def_d} to yield
\bea
&& \left\langle \boldsymbol{f}(\boldsymbol{x}(t_2)) \cdot \boldsymbol{f}(\boldsymbol{x}(t_1)) \right \rangle  =  \int d \boldsymbol{x_1} \int d \boldsymbol{x_2} \, \boldsymbol f(\boldsymbol{x}_2) \cdot \boldsymbol f( \boldsymbol{x}_1) \cr 
&& \times  \int \frac{d \boldsymbol k_1}{(2 \pi)^d} \int \frac{d \boldsymbol k_2}{(2 \pi)^d} \, \exp{[-D \{k_2^2(t_2-t_1)+ k_1^2 t_1\}]} \cr
&& ~~~ \times \exp{[-i \boldsymbol k_2 \cdot (\boldsymbol x_2-\boldsymbol x_1)-i \boldsymbol k_1 \cdot \boldsymbol x_1]}.
\eea
where  $k_i^2= |\boldsymbol{k}_i|^2$. Performing the $\boldsymbol x$ integrals gives
\bea
&&\Big \la \boldsymbol{f}(\boldsymbol{x}(t_2)) \cdot \boldsymbol{f}(\boldsymbol{x}(t_1)) \Big \ra  =  \int \frac{d \boldsymbol k_1}{(2 \pi)^d} \int \frac{d \boldsymbol k_2}{(2 \pi)^d}  \cr
&&\times \tilde {\boldsymbol{f}}(\boldsymbol k_1 - \boldsymbol k_2 ) \cdot \tilde {\boldsymbol{f}}(\boldsymbol k_2)  e^{-D [k_1^2 t_1+k_2^2 (t_2 -t_1)]}
\label{eq:f12_d}
\eea
where $\tilde {\boldsymbol{f}}(\boldsymbol k) = \int d\boldsymbol{z}\, e^{i\boldsymbol k\cdot \boldsymbol z}\boldsymbol f( \boldsymbol z)$ denotes the $d$-dimensional Fourier transform of the force. Substituting  Eq.~\eqref{eq:f12_d} in \eqref{eq:y2_def_d} and performing the integrals over $t_1$ and $t_2$, we get,
\bea
\la y^2(t) \ra &=& \frac 2{D^2} \int \frac{d \boldsymbol k_1}{(2 \pi)^d} \int \frac{d \boldsymbol k_2}{(2 \pi)^d} \frac{\tilde {\boldsymbol{f}}(\boldsymbol k_1 - \boldsymbol k_2 ) \cdot \tilde {\boldsymbol{f}}(\boldsymbol k_2)}{k_1^2 k_2^2(k_1^2-k_2^2)}\cr
&& \times  \left[k_1^2 (1- e ^{-D k_2^2 t})-k_2^2(1- e^{-D k_1^2 t})\right].
\label{eq:y2_kint_d} 
\eea
The $k$-integrals are performed over the $d$-dimensional space. The above expression gives the exact MSD for the process \eqref{eq:yt_approx_d}, but it cannot be evaluated without knowing the explicit form of the force. However, since we are primarily interested in the long-time behavior, we can proceed a bit further. To this end, we consider the Laplace transform of the MSD with respect to time, $M_2(s) = \int_0^\infty dt\, e^{-st} \la y^2(t) \ra$. Performing the Laplace transform of Eq.~\eqref{eq:y2_kint_d} gives
\bea
M_2(s) &=& \frac 2s \int \frac{d\boldsymbol{k}_1}{(2 \pi)^d} \int \frac{d \boldsymbol{k}_2}{(2 \pi)^d} \frac{\tilde {\boldsymbol f}(\boldsymbol{k}_1 - \boldsymbol{k}_2) \cdot \tilde {\boldsymbol f}(\boldsymbol{k}_2)}{(D k_1^2+s)(Dk_2^2+s)}.\qquad \label{eq:y2_laplace_d}
\eea
We are interested in the large $t$ behavior of the MSD, which is controlled by the small $s$ behavior of its Laplace transform $M_2(s)$. The small $s$ behavior of $M_2(s)$ cannot be extracted by naively putting $s=0$ in \eref{eq:y2_laplace_d} since the resulting integral diverges. To extract the leading divergence of this double integral as $s\to 0$, we first split the integral into two 
\bea
M_2(s) = \frac 2s [I_1(s) + I_2(s)],
\eea 
where $I_1(s)$ and $I_2(s)$ denote the contributions from the integral over the region where $(|\boldsymbol k_1|, |\boldsymbol k_2|) \le \Lambda$ with a constant cut-off $\Lambda\sim O(1)$, and  outside this regime where $(\boldsymbol k_1|, |\boldsymbol k_2|)>\Lambda$, respectively.

In the $s \to 0$ limit, $I_2(s)$ converges to a constant. This can be seen by taking the limit in the integrand itself. Therefore
\bea
M_2(s) = \frac 2s I_1(s) + O \left(\frac 1s \right). \label{eq:I1I2}
\eea 
On the other hand, $I_1(s)$ is expected to diverge in the $s \to 0$ limit, and consequently the small $s$ behavior of $M_2(s)$ is determined by the contribution from $I_1(s)$. In the following we extract the small $s$ behavior of $M_2(s)$, by analysing the asymptotic behavior of $I_1(s)$, for short and long-ranged forces, in different dimensions.\\

\noindent {\bf One dimension}.  In one dimension
\begin{equation}
I_1(s) =  \int_{-\Lambda}^{\Lambda} \frac{d{k}_1}{2 \pi} \int_{-\Lambda}^{\Lambda} \frac{d {k}_2}{2 \pi} \frac{\tilde {f}({k}_1 - {k}_2) \cdot \tilde {f}({k}_2)}{(D k_1^2+s)(Dk_2^2+s)}\,. 
\label{eq:I1_1d}
\end{equation}
To extract the small $s$ behavior, we first note that the symmetric nature of the potential ensures that it suffices to consider the only the region $(k_1,k_2)>0$. Using the scaled variables $q_1 = k_1\sqrt{D/s}$ and $q_2 = k_2 \sqrt{D/s}$, we recast Eq.~\eqref{eq:I1_1d} into 
\begin{align}
I_1(s) &= \frac 4{Ds} \int_{0}^{\overline{\Lambda}} \frac{dq_1}{2 \pi} \int_{0}^{\overline{\Lambda}} \frac{dq_2}{2 \pi}\cr
&\times \frac{\tilde f(\sqrt{\frac s D}(q_1 - q_2)) \tilde f(\sqrt{\frac s D} q_2)}{(q_1^2+1)(q_2^2+1)}\,. 
\label{eq:m2s_cutoff}
\end{align} 
Hereinafter we use the shorthand notation  $\overline{\Lambda}\equiv \Lambda \sqrt{D/s}$. 

We now analyze the short-ranged and long-ranged forces separately. For short-ranged forces, the Fourier transform of the force, $\tilde f(k) = -2i \int_0^\infty dz \sin (kz) V'(z)$, is analytic, and to the leading order in $k$, $\tilde f (k) \simeq -ik a_1 $, where $a_1 = 2 \int_0^\infty dz \, z V'(z)$. Therefore
\begin{align}
I_1(s) \simeq \frac {4a_1^2}{D^2} \int_{0}^{\overline{\Lambda}} \frac{dq_1}{2 \pi} \int_{0}^{\overline{\Lambda}} \frac{dq_2}{2 \pi} 
 \frac{q_2(q_2-q_1)}{(q_1^2+1)(q_2^2+1)}. \label{eq:y2s_short1}
\end{align}
The $q_2$-integral is dominated by the contributions from large $q_2$, and hence
\begin{align}
I_1(s) \simeq \frac {4a_1^2}{D^2} \int_{0}^{\overline{\Lambda}} \frac{dq_1}{2 \pi}  \frac 1{q_1^2+1} \int_{0}^{\overline{\Lambda}} \frac{dq_2}{2 \pi} 
 \frac{q_2^2}{q_2^2+1}\,. \label{eq:y2s_short2}
\end{align}
Evaluating the integrals we obtain
\bea
I_1(s) \simeq \frac {a_1^2}{\pi^2 D^2} \tan^{-1}\overline{\Lambda}\left\{\overline{\Lambda} - \tan^{-1}\overline{\Lambda}\right\}. 
\eea
Recalling that $\overline{\Lambda}\equiv \Lambda \sqrt{D/s}$, taking the small $s$ limit, and keeping only the leading small $s$ behavior of $M_2(s)$ we arrive at Eq.~\eqref{eq:y2_smalls_1d}. \\

For long-ranged forces the situation is a bit different since $\tilde f(k)$ cannot be expanded in a Taylor series. The expected small $k$ behavior is $\tilde f(k) \simeq b\, \text{Sgn}(k)|k|^{\alpha}$ where $\alpha >0$ and $b$ is a constant dependent on the specific potential. In that case, we have, from Eq.~\eqref{eq:m2s_cutoff},
\begin{align}
I_1(s) \simeq \frac{4 b^2 s^{\alpha -1}}{D^{1+\alpha}} \int_{0}^{\overline{\Lambda}} \frac{dq_1}{2 \pi}  \frac 1{q_1^2+1} \int_{0}^{\overline{\Lambda}} \frac{dq_2}{2 \pi} 
 \frac{q_2^{2\alpha}}{q_2^2+1}
\end{align}
where we have used the fact that, similar to the short-ranged case, the $q_2$ integral is dominated by the contributions from large $q_2$. Performing the above integrals  and  taking the small $s$ limit yields
\bea
I_1(s) \simeq \frac{b^2 s^{\alpha -1}}{\pi D^{\alpha+1}}[A + B s^{\frac 12 -\alpha} + O(s^{\frac 32-\alpha})],
\eea
where $A$ and $B$ are $s$-independent constants. If $\alpha < 1/2$, we have $I_1(s) \sim s^{\alpha -1}$; for $\alpha > 1/2$, $I_1(s) \sim s^{-1/2}$, similar to the short-ranged scenario. Ignoring the sub-leading contributions we get \eref{eq:msd_s_long} in the main text. \\

\noindent {\bf Two dimensions}. The Fourier transform of the rotational symmetric force reads 
\begin{equation*}
\tilde {\boldsymbol{f}}(\boldsymbol k) = -i \pi \frac{{\boldsymbol k}}{k} \int_0^\infty dz \, z J_1(k z) V'(z)
\end{equation*}
where $J_1(z)$ is the Bessel function of the first kind of order one. Hence, $\tilde {\boldsymbol{f}}(\boldsymbol k) \simeq -ia_2 {\boldsymbol k}$ with $a_2=\frac \pi 2 \int_0^\infty dz\, z^2 V'(z)$ for short-ranged forces in the leading order in $k$. The same arguments as in $d=1$ lead to 
\begin{align}
I_1(s) \simeq \frac{a_2^2s}{D^3} \int_{0}^{\overline{\Lambda}} \frac{dq_1}{2 \pi}  \frac {q_1}{q_1^2+1} \int_{0}^{\overline{\Lambda}} \frac{dq_2}{2 \pi} 
 \frac{q_2^3}{q_2^2+1}.
\end{align}
Performing the integrals, we get
\begin{align*}
I_1(s) \simeq \frac{a_2^2 s}{4 \pi^2 D^3} \log \left(1+\overline{\Lambda}^2\right) \Big[\overline{\Lambda}^2 - \log \left(1+\overline{\Lambda}^2\right)\Big].
\end{align*}
Taking the leading small $s$ behavior of the above expression we get the small $s$ behavior of $M_2(s)$ quoted in Eq.~\eqref{eq:small_s_d2} where we have  put $\Lambda=1$ for simplicity. \\

\noindent {\bf Three dimensions}. In the leading order in $k$, we have $\tilde {\boldsymbol{f}}(\boldsymbol k) \simeq -ia_3 {\boldsymbol k}$ with $a_3 = \frac {4 \pi}3 \int_0^\infty dz \, z^3 V'(z)$. Consequently, the $3d$ version of Eq.~\eqref{eq:y2s_short2} is given by,
\begin{align}
I_1(s) \simeq \frac{a_3^2 s}{(\pi D)^4} \int_{0}^{\overline{\Lambda}} \frac{dq_1}{2 \pi}  \frac {q_1^2}{q_1^2+1} \int_{0}^{\overline{\Lambda}} \frac{dq_2}{2 \pi} 
 \frac{q_2^4}{q_2^2+1}.
\end{align} 
Performing the integrals, taking the small $s$ limit, and using it in  \eref{eq:I1I2}, we get \eref{eq:m2s_3d} in the main text.

\section{Square potential}
\label{app:square_pot}

In this Appendix we provide the details for the computation of the MSD and position distribution for the square-well interaction potential. We start from the backward Feynman-Kac equation in the Laplace space \eqref{eq:Qps_box}, which is a second order homogeneous differential equation except at $x_0= \pm R$. To solve this equation, it is convenient to consider separately three regions and then match the solutions in the different regimes.

\begin{itemize}[leftmargin=*]

\item Region I ($x_0 \ge R$): Here the solution of \eqref{eq:Qps_box} satisfying the boundary condition $\tilde{Q}_p(x_0\to \infty,s)=1/s$ is  
\bea
\tilde{Q}_p(x_0)= \frac 1s + A_1 e^{-x_0\, \sqrt{s/D} } \label{eq:QpsI}
\eea
where $A_1$ is an arbitrary constant to be determined. Hereinafter we usually suppress $s$ variable, e.g., $\tilde{Q}_p(x_0)\equiv \tilde{Q}_p(x_0,s)$. 

\item Region II ($-R\le x_0 \le R$): In this region, the general solution reads,
\bea
\tilde{Q}_p(x_0) = \frac 1s  + A_2 e^{x_0 \, \sqrt{s/D} } + B_2 e^{-x_0 \, \sqrt{s/D}  }  \label{eq:QpsII}
\eea
where $A_2$ and $B_2$ are arbitrary constants.

\item Region III ($x_0 \le -R$): the general solution satisfying the boundary condition $\tilde{Q}_p(x_0) \to -\infty,s)=\frac 1s$
\bea
\tilde{Q}_p(x_0)= \frac 1s + A_3 e^{ x_0 \, \sqrt{s/D}} \label{eq:QpsIII}
\eea
where $A_3$ is an arbitrary constant. 

\end{itemize}

To determine the unknown constants $A_1$, $A_2$, $B_2$ and $A_3$, we use the fact that the solution $\tilde{Q}_p(x_0,s)$ is continuous at $x_0=R$ and $x_0=-R$. Additionally, by integrating Eq.~\eqref{eq:Qps_box} across a small interval around $x=R$ and $x=-R$, we find two jump conditions to be satisfied by the derivatives. Thus we have four conditions:
\begin{eqnarray*}
\tilde{Q}_p(R^+) &=& \tilde{Q}_p(R^-)  \\
 D\partial_{x_0}\! \big[\tilde{Q}_p(R^+)- \tilde{Q}_p(R^-)\big] &=&p V_0 \tilde{Q}_p(R)  \\
\tilde{Q}_p(-R^+) &=& \tilde{Q}_p(-R^-)   \\
D\partial_{x_0}\! \big[\tilde{Q}_p(-R^+)- \tilde{Q}_p(-R^{-})\big] &=& -p V_0 \tilde{Q}_p(-R) 
\end{eqnarray*}
By inserting the explicit solutions \eqref{eq:QpsI}--\eqref{eq:QpsIII} into the above conditions we deduce four linear equations for four unknowns. After simplifications, they read,
\begin{align}
A_1 - A_2 \rho^2 - B_2 &= 0 \label{eq:AB_1} \\
(pV_0 + \sqrt{sD}) A_1 + \sqrt{s D} (\rho^2 A_2 -  B_2) &= - \frac{p V_0}s \rho \label{eq:AB_2} \\
A_2 + B_2 \rho^2 - A_3&=0 \label{eq:AB_3} \\
\sqrt{sD} (A_2 - \rho^2 B_2) + (p V_0 -\sqrt{sD}) A_3 &= - \frac{p V_0}{s}\,\rho. \label{eq:AB_4}
\end{align}
where $\rho=e^{R \sqrt{\frac sD}}$. 
Using Eq.~\eqref{eq:QpsII} we find 
\bea
\mathcal{Q}_p(s)\equiv \tilde{Q}_p(x_0=0,s) = \frac 1s+ A_2+B_2.  \label{eq:Qps_0}
\eea
Solving the linear equations \eqref{eq:AB_1}--\eqref{eq:AB_4} and substituting them in \eqref{eq:Qps_0} gives the result announced in 
Eq.~\eqref{eq:Qp0s_sol}.

To obtain the position distribution in the large time regime, we start from Eq.~\eqref{eq:Qpx0s}. To invert the double Laplace transform, we use the integral representation
\bea
\frac{1}{\sqrt{s}- c_1^2 p^2} \equiv \int_0^{\infty} dN\, \exp{[N (c_1^2 p^2 -\sqrt{s})]} \n 
\eea
and the identity
\bea
\frac 1{\sqrt{4 \pi \sigma^2 N}} \int_{-\infty}^{\infty} dy \, \exp{\Big[- \frac{y^2}{4 \sigma^2 N}- p y\Big]}\equiv e^{p^2 \sigma^2 N} \n
\eea
to rewrite Eq.~\eqref{eq:Qpx0s} as
\begin{align}
\mathcal{Q}_p(s) \simeq \int_{-\infty}^{\infty} dy e^{- p y}  \int_{0}^{\infty} \frac{dN}{\sqrt{4 \pi c_1^2 N s}}\,e^{-N\sqrt{s}- \frac{y^2}{4c_1^2 N}}\,. \label{eq:Qp0s_2}
\end{align}
The Laplace transform $\hat P(y,s) = {\cal L}_{t \to s} P(y,t)$ can then be read off from the above equation, which is given in Eq.~\eqref{eq:P_ys}. Finally, the Laplace transform with respect to $s$ can be inverted using the identity
\bea
{\cal L}^{-1}_{s \to t} \bigg[ \frac{e^{- N\sqrt{s}}}{\sqrt{s}} \bigg]= \frac 1 {\sqrt{\pi t}} \exp\!\bigg(-\frac {N^2}{4t}\bigg).
\eea
Completing these steps we arrive at Eq.~\eqref{eq:Pyt_int}.

\section{Tent potential}
\label{app:Linear_potential}

To solve the backward Fokker-Planck equation \eqref{eq:Qps_diff} for the tent potential \eqref{eq:V_lin} in one dimension, we consider separately four regions: (I) $x_0 \ge R $, (II) $0 < x_0 < R$, (III) $-R < x_0 < 0$ and (IV) $x_0 < -R$. The solutions in neighboring regions can then be matched using the appropriate boundary conditions.

\begin{itemize}[leftmargin=*]
\item $|x_0| > R$: The force vanishes in these regions, and Eq.~\eqref{eq:Qps_diff} reduces to
\bea
 D \partial_{x_0}^2 \tilde Q_p - s \tilde Q_p = -1 \label{eq:R1-4}
\eea
which is solved to yield
\begin{equation*}
\tilde Q_p(x_0) =
\begin{cases}
C_1 e^{x_0\sqrt{s/D}} + C_2 e^{-x_0\sqrt{s/D}} +  \frac 1s & x_0 < -R\\
G_1 e^{x_0\sqrt{s/D}} + G_2 e^{-x_0\sqrt{s/D}} +  \frac 1s & x_0 > R
\end{cases}
\end{equation*}
The boundary conditions \eqref{eq:Qps_bc} imply that $C_2 =0 = G_1$ and therefore
\begin{equation}
\tilde Q_p(x_0) =
\begin{cases}
\displaystyle C_1 e^{x_0\sqrt{s/D}} \, +   \frac 1s & x_0 < -R\\
\displaystyle G_2 e^{-x_0\sqrt{s/D}} +  \frac 1s & x_0 > R
\end{cases}
\end{equation}

\item $0 < x_0 <R$:  In this region, the force is non-zero and  Eq.~\eqref{eq:Qps_diff} reduces to
\bea
  \partial_{x_0}^2 \tilde Q_p - \lambda_+^2 \tilde Q_p =-1, \quad \lambda_+=\sqrt{\frac{s + pV_0/R}{D}}
\eea
 which has the general solution
 \bea
 \tilde Q_p(x_0) = \frac{1}{D\lambda_+^2 } + A_1 e^{x_0\lambda_+}+A_2 e^{-x_0\lambda_+}\,.
 \eea

\item $-R < x_0 <0$: In this region, Eq.~\eqref{eq:Qps_diff} reduces to
\bea
\partial_{x_0}^2 \tilde Q_p -  \lambda_-^2 \tilde Q_p =-1, \quad \lambda_-=\sqrt{\frac{s - pV_0/R}{D}}
\eea
which has the general solution,
\bea
 \tilde Q_p(x_0) = \frac{1}{D\lambda_-^2 } + B_1 e^{x_0\lambda_-}+B_2 e^{-x_0\lambda_-}\,.
\eea
\end{itemize}

To determine the $s$-dependent constants $A_{1,2}$, $B_{1,2}$, $C_1$ and $G_2$  we note that $\tilde Q_p(x_0)\equiv \tilde Q_p(x_0,s)$ must be a continuous function of $x_0$. Moreover, by integrating  \eqref{eq:Qps_diff} for the  $f(x)$ given by \eqref{eq:fx_linear} across $x_0=\pm R$ and  $x_0 = 0$, we find that $\partial_{x_0} \tilde Q_p(x_0)$ must also be continuous. Thus
\begin{eqnarray*}
\tilde Q_p(0^+) &=&  \tilde Q_p(0^-), ~~~~~\partial_{x_0}\tilde Q_p(0^+) =  \partial_{x_0} \tilde Q_p(0^-), \\
\tilde Q_p(R^+) &=&  \tilde Q_p(R^-), ~~~~\partial_{x_0}\tilde Q_p(R^+) =  \partial_{x_0} \tilde Q_p(R^-), \\
\tilde Q_p(-R^+) &=&  \tilde Q_p(-R^-), ~~\partial_{x_0}\tilde Q_p(-R^+) = \partial_{x_0} \tilde Q_p(-R^-).
 \end{eqnarray*}
Inserting the above solutions into these conditions we get six linear equations. We are particularly interested in the behavior of $\mathcal{Q}_p(s)\equiv \tilde Q_p(0,s)$, so we eliminate $C_1$ and $G_2$ leaving us with two long equations 
\begin{eqnarray*}
A_1 \left(1 + \lambda_+\sqrt{\tfrac{D}{s}} \right) e^{\lambda_+ R} &+& A_2 \left(1 - \lambda_+\sqrt{\tfrac{D}{s}}  \right) e^{- \lambda_+ R} \\
&=& \frac{p V_0/R}{s(s+ p V_0/R)}, \\
B_1 \left(1 - \lambda_-\sqrt{\tfrac{D}{s}} \right) e^{-\lambda_- R} &+& B_2 \left(1 + \lambda_-\sqrt{\tfrac{D}{s}}  \right) e^{- \lambda_+ R} \\
&=& -\frac{p V_0/R}{s(s+ p V_0/R)} 
\end{eqnarray*}
and two shorter equations
\begin{equation*}
\begin{split}
A_1 + A_2 - B_1 -B_2 &= \frac{2 p V_0/R}{s^2 - p^2 V_0^2/R^2}, \\
A_1 - A_2 &= \frac{\lambda_-}{\lambda_+}(B_1 - B_2).
\end{split}
\end{equation*}

These linear equations can be readily solved, but we omit long explicit formulae for $A_{1,2}$ and $B_{1,2}$. We are primarily interested in the long-time and large $y$ behavior, so it suffices to consider the small $s$ and small $p$ limiting behavior of $\mathcal{Q}_p(s)$. As before, we first compute the second moment. Expanding $\mathcal{Q}_p(s)$ around $p=0$ and taking the coefficient of $p^2$ we obtain
${\cal L}_{t \to s} \la y^2(t) \ra \simeq \frac{2 R V_0^2}{3} (D s)^{-\frac{3}{2}}$. Performing the inverse Laplace transform we arrive at the announced result \eqref{eq:y2t_lin}.  

To compute the full distribution $P(y,t)$, we consider $\mathcal{Q}_p(s)$, keep the leading order contributions in the limit of $s \to 0$ and $p \to 0$, and deduce the announced Eq.~\eqref{eq:Qps_lin}  which leads to the scaling form of the distribution.

\bibliography{references-active}

\end{document}